\documentclass[11pt, a4paper]{article}
\pdfoutput=1
\usepackage{jcappub}
\usepackage[utf8]{inputenc}
\usepackage[english]{babel}
\usepackage{amsmath, amssymb}
\usepackage{graphicx}
\usepackage{hyperref}

\usepackage{xcolor}
\usepackage[normalem]{ulem}

\newcommand{\e}{\mathrm{e}}
\newcommand{\im}{\mathrm{i}}

\makeatletter
\newcommand*{\dd}{\mathop{}\!{\operator@font d}}
\makeatother

\newcommand{\mH}{\mathcal{H}}
\newcommand{\vk}{\mathbf{k}}
\newcommand{\vq}{\mathbf{q}}
\newcommand{\vp}{\mathbf{p}}
\newcommand{\vv}{\mathbf{v}}

\newcommand{\vnhat}{\mathbf{\hat{n}}}

\newcommand{\zmatch}{z_{\mathrm{match}}}
\newcommand{\etaD}{\eta}
\newcommand{\etaDmatch}{\etaD_{\mathrm{match}}}
\newcommand{\etaDasymp}{\etaD_{\mathrm{asymp}}}
\newcommand{\etaDini}{\etaD_{\mathrm{ini}}}
\newcommand{\mnu}{m_{\nu}}
\newcommand{\fnu}{f_{\nu}}
\newcommand{\omegaoverfsq}{\frac{\Omega_m}{f^2} }

\newcommand{\alphas}{\alpha^{\mathrm{s}}}

\newcommand{\nr}{\mathrm{nr}}
\newcommand{\cb}{\mathrm{cb}}
\newcommand{\cdm}{\mathrm{cdm}}
\newcommand{\mm}{\mathrm{m},\mathrm{m}}
\newcommand{\lin}{\mathrm{lin.}}
\newcommand{\oneloop}{\text{1-loop}}
\newcommand{\twoloop}{\text{2-loop}}
\newcommand{\eV}{\mathrm{eV}}
\newcommand{\Mpc}{\mathrm{Mpc}}
\newcommand{\LCDM}{\Lambda\mathrm{CDM}}
\newcommand{\wCDM}{w\mathrm{CDM}}

\newcommand{\bea}{\begin{eqnarray}}
\newcommand{\eea}{\end{eqnarray}}
\newcommand{\be}{\begin{equation}}
\newcommand{\ee}{\end{equation}}
\newcommand{\nn}{\nonumber}

\subheader{\small
    \vspace*{-1.28cm}
    \begin{flushright}
        TUM-HEP-1273/20
    \end{flushright}
}

\title{Loop corrections to the power spectrum for massive neutrino cosmologies with full time- and scale-dependence}

\author{Mathias Garny,}
\author{Petter Taule}

\affiliation{Physik Department T31,\\
    James-Franck-Str. 1, Technische Universit\"at M\"unchen,\\
D-85748 Garching, Germany}

\emailAdd{mathias.garny@tum.de}
\emailAdd{petter.taule@tum.de}

\abstract{%
    Loop corrections to the power spectrum are frequently computed using
    approximate non-linear kernels adopted from an Einstein de-Sitter (EdS)
    cosmology.  We present an algorithm that allows us to take the full time-
    and scale-dependence of the underlying fluid dynamics into account, and
    apply it to assess the impact of neutrino free-streaming on the 1- and
    2-loop matter power spectrum. Neutrino perturbations are described by a
    hybrid Boltzmann/two-fluid model, that we show to agree with the CLASS
    Boltzmann solution at the linear level when using an appropriate effective
    neutrino sound velocity. We then use this scheme at 1- and 2-loop to
    perform a precision comparison of the matter power spectrum with simplified
    treatments of massive neutrinos. For a commonly adopted scheme using cold
    dark matter+baryon perturbations and EdS kernels to compute non-linear
    corrections, we find deviations above $1\%$ for $k>0.15~h/\Mpc$ at $z=0$, and
    sub-percent agreement on weakly non-linear scales at $z=0.5$. We also
    demonstrate the impact of the EdS approximation on the 2-loop power spectrum
    for massless neutrinos.
}

\begin{document}

\maketitle

\section{Introduction}

Mapping out the large-scale structure (LSS) of the Universe is one of the key
advances that is driving precision cosmology, and current
\cite{Abbott:2018xao,Alam:2020sor,Bocquet:2018ukq} and near future
\cite{Amendola:2016saw,Levi:2019ggs,Ellis:2012rn,Abell:2009aa}
surveys will allow us to potentially detect even small deviations from
the $\Lambda$CDM model at the (sub-)percent level. Apart from testing possible explanations of the
apparent tension in the value of the Hubble constant $H_0$ within extended
cosmological models, precise LSS observations enable us to scrutinize
the properties of dark energy or probe tiny deviations from the
cold dark matter hypothesis on large scales~\cite{Amendola:2016saw,Bechtol:2019acd,Gluscevic:2019yal}.
Indeed, even within the standard cosmological model, the non-baryonic matter is expected to be composed of
a mixture of cold and hot components, the latter being provided by the
cosmic neutrino background. Identifying its impact on structure formation
offers the prospect to infer the absolute neutrino mass scale within the
$\Lambda$CDM framework~\cite{Audren:2012vy,Amendola:2016saw,Boyle:2017lzt,Brinckmann:2018owf,Mishra-Sharma:2018ykh,Chudaykin:2019ock,Xu:2020fyg}.

A major technical challenge when exploiting LSS data is to provide precise and robust
theoretical predictions for a large set of cosmological parameters, as required in
a statistical analysis.
Given the extended coverage of weakly non-linear scales in
future surveys, major efforts have been devoted to obtain a suitable perturbative description~\cite{Bernardeau:2001qr}.
Useful ingredients that have been explored in this context include the perturbative bias expansion~\cite{Desjacques:2016bnm},
models of redshift-space distortions (RSD)~\cite{Scoccimarro:1999ed},
analyses of stress-tensor contributions~\cite{Pueblas:2008uv},
effective field theory methods~\cite{Baumann:2010tm}, as well as IR resummation capturing the non-linear broadening
of the baryon acoustic oscillation (BAO) feature in the two-point correlation
function~\cite{Crocce:2005xy,Senatore:2014via,Baldauf:2015xfa,Blas:2016sfa}.
A viable analysis strategy consists in capturing bias, RSD as well as the impact of small (UV) scales
by a number of free parameters, that can be marginalized over.
Approaches along these lines have, in various implementations that differ in the
underlying model but yield consistent results, been applied to analyze the full shape of
BOSS galaxy clustering data~\cite{Sanchez:2016sas,DAmico:2019fhj,Ivanov:2019pdj,Troster:2019ean}.

To achieve a theoretical uncertainty at or below the percent level, every element of the
perturbative description needs to be scrutinized. In this work we focus on the impact of
time- and scale-dependence of the growth of structures on the computation of non-linear
corrections to the power spectrum. While the framework that we present here is applicable
to a large variety of extended cosmological models, we discuss in particular the case of massive neutrinos.
Within the framework of standard perturbation theory (SPT), the
density contrast is expanded perturbatively around the linear solution,
yielding wavenumber-dependent kernels $F_n(\vk_1,\dots,\vk_n)$ at the $n$-th order. In turn, these
kernels furnish the perturbative expansion of the power spectrum. A major simplification
occurs for an Einstein de-Sitter (EdS) Universe that is dominated by matter. In this case the time-dependence
of the kernels factorizes, and they can be computed using well-known algebraic recursion relations.
This simplification is in general not possible in any realistic cosmological model, including $\Lambda$CDM.
Nevertheless, EdS-SPT kernels are widely used also for $\Lambda$CDM, after correcting
for the main impact of dark energy by using a suitably rescaled time variable~\cite{Bernardeau:2001qr}.

When including massive neutrinos, the linear growth rate becomes scale-dependent due to neutrino
free streaming~\cite{Lesgourgues:2006nd}. An accurate description at the linear level requires to solve Boltzmann equations
for the neutrino perturbations, as is done routinely in codes such as CLASS~\cite{Blas:2011rf} or CAMB~\cite{Howlett:2012mh}.
However, this quickly becomes impractical beyond linear perturbation theory~\cite{Fuhrer:2014zka}. Fortunately, for values of the neutrino mass
targeted by future surveys, their distribution becomes non-relativistic long before non-linear
corrections become relevant. This enables a hybrid description based on the full Boltzmann
evolution at high redshifts, and an effective two-fluid description at low redshift, that is amenable
to perturbation theory~\cite{Blas:2014}. In this work we present an improved version of the two-fluid description
that reproduces the linear evolution predicted by the Boltzmann equation at the permille level
for cold dark matter, and with better than percent accuracy for neutrino perturbations. Given
that the latter constitute only a small fraction of the total density contrast, this translates
into a negligible error for the matter power spectrum. Free-streaming is captured by an effective,
scale-dependent sound velocity within the fluid description.
Based on this two-fluid model, we obtain
results for the 1- and 2-loop power spectrum, taking the exact time- and scale-dependence of the
underlying fluid dynamics into account. We compare these results to various simplified approximation
schemes for massive neutrinos that are commonly used in this context, and assess in how far they
capture the impact of free-streaming.
Our work should be considered as a starting point, aiming at an inclusion of the effects
of redshift space distortions, stress-tensor/effective field theory corrections as well as relativistic
effects~\cite{Tram:2018znz} in the future.
In addition, the framework for computing non-linear corrections employed in this work is rather versatile and
can be applied to extended cosmological models with scale-dependent growth in the future.
Aiming at an accurate and efficient description within the weakly non-linear regime, the perturbative approach
is complementary to $N$-body simulations \cite{Bird:2011rb,Dakin:2017idt,Castorina:2015bma}
and (halo-)models \cite{Hannestad:2020rzl} investigating the effect of massive neutrinos on smaller scales.

In Sec.\,\ref{sec:num_setup}, we introduce the general setup for computing loop corrections to the
power spectrum taking exact time- and scale-dependence into account, keeping the discussion as general as possible.
A first application to $\Lambda$CDM is discussed in Sec.\,\ref{sec:eds_approx}, where we assess the error of the EdS-SPT approximation at 2-loop.
We turn to massive neutrinos in Sec.\,\ref{sec:2fluid_lin}, where we introduce the hybrid Boltzmann/two-fluid description and quantify
its accuracy at the linear level. In Sec.\,\ref{sec:2fluid_nonlin}, we present results for the 1- and 2-loop power spectrum for massive
neutrinos and compare them to various approximate schemes, before concluding in Sec.\,\ref{sec:conclusions}. In the appendices we
discuss the validation and convergence checks of our numerical algorithm.

\section{Evaluation of loop corrections with exact time- and scale-dependence}
\label{sec:num_setup}

In this section we discuss the strategy for computing loop corrections to the power
spectrum with exact time- and scale-dependence that allows us to go up to 2-loop.
After briefly reviewing the standard formalism and setting up our notations, we
introduce a generic extension of standard perturbation theory
that can capture a large variety of cosmological models beyond $\LCDM$,
as well as effective descriptions of structure formation. We then outline the
algorithm for computing non-linear corrections to the power spectrum within
this framework, leaving some details relevant for the application to massive
neutrinos to Sec.\,\ref{sec:2fluid_nonlin}.

\subsection{Standard Perturbation Theory}

We start by briefly reviewing the formalism for standard perturbation theory, and introducing some basic notations.
The equations of motion for the density contrast $\delta$ and velocity
divergence $\theta \equiv \partial_i \vv^{i}$ (neglecting vorticity) in Fourier
space are
\begin{align}
    \partial_{\tau} \delta(\vk,\tau) + \theta(\vk,\tau) & =
        - \int_{\vk_1,\vk_2}\delta_D (\vk - \vk_{12}) \alpha(\vk_1,\vk_2)
        \theta(\vk_1, \tau) \delta(\vk_2, \tau) \, ,
    \nonumber \\
    \partial_{\tau} \theta(\vk,\tau) + \mH \theta(\vk, \tau)
    + \frac{3}{2} \mH^2 \Omega_m \delta(\vk,\tau) & =
        - \int_{\vk_1,\vk_2}\delta_D (\vk - \vk_{12}) \beta(\vk_1,\vk_2)
        \theta(\vk_1, \tau) \theta(\vk_2, \tau) \, ,
    \label{eq:cont_euler}
\end{align}
where $\tau$ is conformal time, $\mH = \dd \ln a/\dd \tau$, $\Omega_m$ the
time-dependent matter density parameter, $\delta_D$ denotes the Dirac delta
function and we use the shorthand notations $\vk_{12} = \vk_1 + \vk_2$ and
$\int_{\vk} = \int \dd^3 \vk$. The mode coupling functions are
\begin{equation}
    \alpha(\vk_1,\vk_2) = 1 + \frac{\vk_1\cdot\vk_2}{k_1^2}\, ,\quad
    \beta(\vk_1,\vk_2) = \frac{(\vk_1 + \vk_2)^2 (\vk_1 \cdot \vk_2)}{2 k_1^2 k_2^2}\, .
\end{equation}
Introducing the tuple $\psi_a = (\delta, -\theta/\mH f)$, the above equations may be written in the compact form
\begin{equation}
    \partial_{\etaD} \psi_a(\vk,\etaD) + \Omega_{ab}(\etaD) \psi_b(\vk,\etaD) =
    \int_{\vk_1, \vk_2} \delta_D(\vk-\vk_{12})\gamma_{abc}(\vk,\vk_1,\vk_2)
        \psi_b(\vk_1,\etaD) \psi_c(\vk_2,\etaD)\, ,
    \label{eq:eom_compact}
\end{equation}
where $\etaD = \ln D$ and $f = \dd \ln D/\dd \ln a$, $D$ being the linear
growth factor. The matrix $\Omega_{ab}$ is given by
\begin{equation}
    \Omega_{ab}(\etaD) =
    \begin{pmatrix}
        0 & -1 \\
        - \frac{3}{2} \omegaoverfsq & \frac{3}{2} \omegaoverfsq - 1
    \end{pmatrix}\,,
    \label{eq:1fluid_omega_mat}
\end{equation}
and the only non-zero elements of $\gamma_{abc}$ are
\begin{equation}\label{eq:gamma_lcdm}
        \gamma_{121}(\vk,\vk_1,\vk_2) = \alpha(\vk_1,\vk_2)\,,\quad
        \gamma_{222}(\vk,\vk_1,\vk_2) = \beta(\vk_1,\vk_2)\, .
\end{equation}

In an EdS Universe, $\Omega_m = 1$ and $f = 1$, which makes $\Omega_{ab}$
time-independent. Even though $\Omega_m$ and $f$ differ significantly from
unity entering vacuum energy domination for a $\LCDM$ or $\wCDM$ cosmology, the
ratio $\Omega_m/f^2$ remains close to 1. In addition, at the linear level, the
deviation of $\Omega_m/f^2$ from unity does not affect the growing mode (with
eigenvector proportional to $(1,1)$). This is usually taken as an argument
to replace the matrix $\Omega_{ab}$ by
\begin{equation}\label{eq:OmEdS}
    \Omega^{\rm EdS}_{ab}(\etaD) =
    \begin{pmatrix}
        0 & -1 \\
        - \frac{3}{2}  & \frac{1}{2}
    \end{pmatrix}\,,
\end{equation}
even when considering $\LCDM$ cosmology, leading to the conventional form of SPT, that we denote by
EdS-SPT. The virtue of this approximation is that, when solving \eqref{eq:eom_compact} perturbatively
in the linear density contrast $\delta^{(1)}(\vk,\etaD)=\e^{\eta}\delta_0(\vk)$, the time-dependence
factorizes,
\begin{equation}\label{eq:Fn}
    \delta(\vk, \etaD) = \sum_{n=1}^{\infty}
        \int_{\vq_1,\dotsc,\vq_n} \delta_D(\vk - \vq_{1\cdots n}) \,
        \e^{n\etaD}\,F_n (\vq_1,\dotsc,\vq_n) \,
        \delta_0(\vq_1)\dotsb \delta_0(\vq_n)\, ,
\end{equation}
where $\vq_{1\cdots n}=\sum_i\vq_i$. A similar expansion holds for $\psi_2=-\theta/\mH f$
with kernels $G_n$. The EdS-SPT kernels can be obtained from well-known algebraic
recursion relations \cite{Bernardeau:2001qr}, and furnish the perturbative expansion of the power spectrum
\be\label{eq:Pdd}
  \langle\delta(\vk,\etaD)\delta(\vk',\etaD)\rangle=\delta_D(\vk+\vk')P(k,\etaD)\,.
\ee
The EdS-SPT kernels are frequently used also for models beyond $\LCDM$, and in
particular for massive neutrinos. In this work we scrutinize this approximation
using the framework described below.

\subsection{Extension beyond SPT and numerical evaluation}
\label{sec:extension}

Going beyond EdS-SPT generically leads to a breakdown of the factorization of
time- and scale-dependence when expanding the density contrast perturbatively.
This poses a significant technical increase in complexity, and has
therefore been little explored so far beyond 1-loop. In this section we
present the setup that we employ to tackle this challenge.
The algorithm is an extension of the one employed in \cite{Blas:2013bpa,Blas:2013aba,Blas:2015tla}.

We start by introducing a generic extension of SPT that potentially encompasses many models beyond $\LCDM$ as well
as effective descriptions of structure formation. Below we will apply this setup
to massive neutrino cosmologies, and also assess the accuracy of the EdS approximation
for $\Omega_{ab}$ within $\LCDM$.
We assume that cosmic perturbations can be
described by a set of fields $\psi_a(\vk,\etaD)$, allowing for multiple species. For an $N$-fluid
model the vector $\psi_a$ contains the density contrast $\delta_i$ and the
(suitably normalized) divergence of the peculiar
velocity field $\theta_i$ for each fluid component $i$, and the index $a$ runs from $1$ to $2N$.
The full set of auto- and cross power spectra of these perturbations is given by
\be\label{eq:Pab}
  \langle\psi_a(\vk,\etaD)\psi_b(\vk',\etaD)\rangle=\delta_D(\vk+\vk')P_{ab}(k,\etaD)\,.
\ee
We further assume that the non-linear equation of motion
of $\psi_a$ can be brought into a form analogous to Eq.\,\eqref{eq:eom_compact},
\begin{equation}
    \partial_{\etaD} \psi_a(\vk,\etaD) + \Omega_{ab}(k,\etaD) \psi_b(\vk,\etaD) =
    \int_{\vk_1, \vk_2} \delta_D(\vk-\vk_{12})\gamma_{abc}(\vk,\vk_1,\vk_2) \psi_b(\vk_1,\etaD) \psi_c(\vk_2,\etaD)\,.
    \label{eq:eom_compact_general}
\end{equation}
The left-hand side comprises the linear
equation of motion, being captured by a $2N\times 2N$ matrix $\Omega_{ab}(k,\etaD)$. We allow
for both a time- and wavenumber-dependence, with $k\equiv|\vk|$. The dependence on wavenumber occurs for
example when taking an effective viscosity and sound velocity into account for dark matter \cite{Blas:2015tla},
and is also important for massive neutrinos \cite{Blas:2014}. In addition,  $\gamma_{abc}(\vk,\vk_1,\vk_2)$ denote
the non-linear vertices, that capture quadratic non-linearities in the
equations of motion in real space. For $\LCDM$ they are given by Eq.\,\eqref{eq:gamma_lcdm}, but may also include
additional non-linear terms in general.

The equation of motion needs to be complemented by suitable initial conditions.
We assume that the initial conditions for all perturbation variables are correlated, as is the
case for the familiar adiabatic initial conditions. Consequently, they can be brought into the form
\be
  \psi_a(\vk,\etaDini)= u_a(k,\etaDini)\delta_0(\vk)\, ,
\ee
where $\etaDini$ should be chosen sufficiently late
after recombination, but long before the onset of non-linear evolution on scales relevant for large-scale
structure formation. Neglecting primordial non-Gaussianity,
 $\delta_0(\vk)$ can be taken as a Gaussian random field, of which we ultimately only need to
know the (linear) power spectrum $\langle\delta_0(\vk)\delta_0(\vk')\rangle=\delta_D(\vk+\vk')P_0(k)$ as an input for the
computation of non-linear corrections.
The vector $u_a(k,\etaDini)$ determines the relative normalization of the initial values for the various
perturbation variables. For a $\LCDM$ cosmology featuring only a single fluid component, it can be chosen to agree
with the growing mode eigenvector $u_a(k,\etaDini)=(1,1)$ with $\etaDini$ deep inside the matter dominated era.
We refer to the following sections for some details on how the initial conditions are implemented for
the respective applications.

The perturbative expansion of the solution to Eq.\,\eqref{eq:eom_compact_general} can be written in the form
\begin{equation}\label{eq:Fan}
    \psi_a(\vk, \etaD) = \sum_{n=1}^{\infty}
        \int_{\vq_1,\dotsc,\vq_n} \delta_D(\vk - \vq_{1\cdots n}) \,
        F_a^{(n)} (\vq_1,\dotsc,\vq_n;\etaD) \,
        \delta_0(\vq_1)\dotsb \delta_0(\vq_n)\, .
\end{equation}
Importantly, compared to the EdS-SPT case given in Eq.\,\eqref{eq:Fn}, the time-dependence does
in general \emph{not} factorize. To obtain an evolution equation for the kernels, we insert
Eq.\,\eqref{eq:Fan} into the equation of motion \eqref{eq:eom_compact_general} and
collect terms of equal order in $\delta_0$. This yields an equation of motion for the kernels
\bea
    \lefteqn{
        \partial_{\etaD}F_a^{(n)}(\vq_1,\dotsc,\vq_n;\etaD) + \Omega_{ab}(k,\etaD)
    F_b^{(n)}(\vq_1,\dotsc,\vq_n;\etaD) }\nn\\
    & = &
    \sum_{m=1}^{n-1} \Big[\gamma_{abc}(\vk,\vq_{1\dotsb m},\vq_{m+1\dotsb n})
    F_b^{(m)}(\vq_1,\dotsc,\vq_m;\etaD)
    F_c^{(n-m)}(\vq_{m+1},\dotsc,\vq_n;\etaD)\Big]_{\rm sym.}
    \, .\quad
    \label{eq:eom_kernels}
\eea
Here the right hand side is understood to be symmetrized with respect to $N_m=\frac{n!}{m!(n-m)!}$
permutations exchanging momenta in the $\{\vq_1,\dotsc,\vq_m\}$ set with momenta
in the $\{\vq_{m+1},\dotsc,\vq_n\}$ set (meaning the sum of all permuted expressions divided by $N_m$).
Furthermore, $\vk = \sum_i \vq_i$.

For any given set of wavevectors, Eq.\,\eqref{eq:eom_kernels} is a set of coupled ordinary differential equation for the $2N$ kernels
$F_a^{(n)}(\vq_1,\dotsc,\vq_n;\etaD)$ with $a=1,\dots,2N$. One can easily check that for $N=1$,
when replacing $\Omega_{ab}(k,\etaD)$ by the constant
EdS matrix given in Eq.\,\eqref{eq:OmEdS}, along with the vertices from Eq.\,\eqref{eq:gamma_lcdm},
one recovers the usual EdS-SPT recursion relations for the kernels in the limit $\etaDini\to-\infty$, with a factorized time-dependence
$F_1^{(n)}\to \e^{n\etaD}F_n$ and $F_2^{(n)}\to \e^{n\etaD}G_n$.

Inserting the perturbative expansion Eq.\,\eqref{eq:Fan} into Eq.\,\eqref{eq:Pab} and using the
Wick theorem yields the ``loop'' expansion of the power spectrum
\be
  P_{ab}(k,\etaD)=P_{ab}^{\rm lin}(k,\etaD)+P_{ab}^{\oneloop}(k,\etaD)+P_{ab}^{\twoloop}(k,\etaD)+\dots\,,
\ee
with
\bea
  P_{ab}^{\rm lin}(k,\etaD) &=& F_a^{(1)}(k;\etaD)P_0(k)F_b^{(1)}(k;\etaD)\,,\nn\\
  P_{ab}^{\oneloop}(k,\etaD) &=& \int_\vq P_0(q) \Big[ 3F_a^{(1)}(k;\etaD)P_0(k) F_b^{(3)}(\vk,\vq,-\vq;\etaD)\nn\\
  && {} + 3F_a^{(3)}(\vk,\vq,-\vq;\etaD) P_0(k) F_b^{(1)}(k;\etaD)\nn\\
  && {} + 2F_a^{(2)}(\vk-\vq,\vq;\etaD) P_0(|\vk-\vq|) F_b^{(2)}(\vk-\vq,\vq;\etaD) \Big]\,, \nn\\
  P_{ab}^{\twoloop}(k,\etaD) &=& \int_{\vq,\vp} P_0(q)P_0(p) \Big[ 15F_a^{(1)}(k;\etaD)P_0(k) F_b^{(5)}(\vk,\vp,-\vp,\vq,-\vq;\etaD)\nn\\
  && {} + 15F_a^{(5)}(\vk,\vp,-\vp,\vq,-\vq;\etaD)P_0(k)F_b^{(1)}(k;\etaD) \nn\\
  && {} + 12F_a^{(2)}(\vk-\vq,\vq;\etaD) P_0(|\vk-\vq|) F_b^{(4)}(\vk-\vq,\vq,\vp,-\vp;\etaD) \nn\\
  && {} + 12F_a^{(4)}(\vk-\vq,\vq,\vp,-\vp;\etaD) P_0(|\vk-\vq|) F_b^{(2)}(\vk-\vq,\vq;\etaD) \nn\\
  && {} + 9F_a^{(3)}(\vk,\vq,-\vq;\etaD)P_0(k) F_b^{(3)}(\vk,\vp,-\vp;\etaD) \nn\\
  && {} + 6F_a^{(3)}(\vk-\vp-\vq,\vp,\vq;\etaD)P_0(|\vk-\vp-\vq|) F_b^{(3)}(\vk-\vp-\vq,\vp,\vq;\etaD) \Big]\,. \nn\\
\eea
These expressions are analogous to the familiar SPT loop integrals, except that here the generalized time-dependent
kernels $F_a^{(n)}$ appear. Note that the indices $a,b$ label the components of $\psi_a$, such that for example $P_{11}$
denotes the density auto power spectrum, $P_{22}$ the velocity divergence spectrum, and $P_{12}$ the cross spectrum. In
case of multiple fluids, $P_{13}$ is the cross power spectrum between the density contrast of the first and second component, etc.
We stress that these indices are unrelated to the conventional naming of the various summands in the 1- and 2-loop integrals
within SPT according to the perturbative order of the kernels. Furthermore, we emphasize that, in general, even the linear
evolution can be non-trivial, being captured by the transfer functions $F_a^{(1)}(k;\etaD)$.

For numerical evaluation, we further modify the expressions inside the square brackets to ensure that large cancellations that
occur when one of the arguments of $P_0$ goes to zero are accounted for already at the integrand level \cite{Blas:2013bpa,Blas:2013aba,Carrasco:2013sva}.
We then compute the loop integrals using the \texttt{Suave} routine of the Monte-Carlo integration package CUBA \cite{Hahn:2004fe}.
The numerical algorithm at 2-loop is an extended version compared to \cite{Blas:2013bpa,Blas:2013aba,Blas:2015tla}, and can be summarized as follows:
\begin{enumerate}
\item The Monte-Carlo loop integrator calls the routine computing the integrand for a given set of wavenumbers $\vp,\vq$. The external
    wavevector is fixed to $\vk=(0,0,k)$ (without loss of generality).
\item All vertices $\gamma_{abc}(\vk_{12},\vk_1,\vk_2)$ are computed and stored for all sets of possible wavevectors $\vk_{1}$ and $\vk_{2}$ of the form
\be
  \epsilon_0\vk+\epsilon_1\vp+\epsilon_2\vq,\qquad \epsilon_0=0,1,\;\epsilon_{1,2}=0,\pm 1\,.
\ee
\item For each kernel $F_a^{(n)}$, evaluated with the particular set of wavevectors occurring in the loop integrand, we
solve the set of ordinary differential equations \eqref{eq:eom_kernels} \emph{numerically}. This requires to solve also for
the lower order kernels $F_a^{(m)}$ and $F_a^{(n-m)}$ with $m=1,\dots,n-1$ evaluated on a subset of wavevectors, which
can be done recursively. We temporarily store the time-dependent kernels on a sufficiently dense grid between $\etaDini$ and the
output time. This leads to a significant reduction in computing time due to the recursive structure of Eq.\,\eqref{eq:eom_kernels} and
since only a limited number of combinations of wavevectors can occur as an argument.
We describe the implementation of initial conditions for the kernels in the sections below.
\end{enumerate}

Let us finally comment on an alternative scheme that can capture a non-trivial time-dependence, known as time-dependent renormalization
group (TRG) \cite{Pietroni:2008jx}, which is based on solving coupled equations for the power- and bispectrum. As shown in \cite{Audren:2011ne},
this scheme can be considered being equivalent to a 1-loop computation of the power spectrum that allows to take an arbitrary time-dependence
of $\Omega_{ab}$ into account. Extending this formalism to 2-loop would require to include also the trispectrum, which represents significant
technical challenges \cite{Juergens:2012ap}. In addition, even when neglecting the trispectrum, the practical feasibility of the
algorithm depends on being able to reduce the computation to a certain moment of the bispectrum.
However, this is no longer possible when $\Omega_{ab}$ depends also on wavenumber, as is the case when including an effective
viscosity and sound velocity, and also for massive neutrinos. In addition, for an $N$-fluid scheme, the TRG approach would require to
track the time-dependence of all $2N(2N-1)$ power spectra, while for the approach pursued here it is sufficient to compute the $2N$ kernels $F_a^{(n)}$.
Note that the algorithm presented here can be extended to the 1-loop bispectrum with exact time- and scale-dependence \cite{Floerchinger:2019eoj}.

\section{Validity of EdS approximation}
\label{sec:eds_approx}

In this section we  turn to our first
application: relaxing the commonly used EdS approximation and computing the
matter power spectrum for the first time at 2-loop from kernels with exact
time-dependence. The effect of the EdS approximation on the 1-loop correction
to the power spectrum has been studied previously in the
literature~\cite{Bernardeau:1993qu,Takahashi:2008yk,Sefusatti:2011cm,Fasiello:2016qpn,Donath:2020abv},
where analytic expressions for the relevant time-dependent kernels in
generalized cosmologies have been derived. It has been shown that the 1-loop term
in the EdS approximation is accurate to less than 1\% at $z = 0$ in the mildly
non-linear regime. We find that adding the 2-loop correction, the inaccuracy
increases to more than 1\% at $k = 0.2~h/\Mpc$.
The approximation works better at earlier times, and indeed we find only a few
permille deviation on the relevant scales at $z = 0.5$. Future low-redshift
surveys may achieve a sensitivity close to this mark,
hence it is worth studying the quantitative effects of the EdS approximation on
cosmological observables.

Our analysis is based on a $\LCDM$ cosmology with $h = 0.6756$,
$\Omega_{\mathrm{b}} = 0.04828$, $\Omega_{\cdm} = 0.2638$, $n_s = 0.9619$ and
$A_s = 2.215\cdot 10^{-9}$. The input linear power spectrum as well as the ratio
$\Omega_m/f^2$ is taken from the Boltzmann solver CLASS.
As described above, to compute the 1- and 2-loop power spectrum,
we solve Eq.\,\eqref{eq:eom_kernels} numerically for
each kernel and set of wavenumbers needed in the Monte-Carlo integration.
We initialize the time evolution deep inside the matter dominated era
(specifically, $z_{\mathrm{ini}} = 25$). Since $\Omega_m/f^2$ is very close to unity
at that time, it is possible to use the conventional EdS-SPT kernels as \emph{initial conditions}.
In Appendix~\ref{app:ana_kernels}, we compare our
numerical 1-loop result with the analytic time-dependent kernels from
Ref.~\cite{Bernardeau:1993qu}, finding agreement at the sub-permille level and within the numerical error
bar of the Monte Carlo integration. We performed numerous additional convergence checks of the
numerical implementation that are discussed in Appendix~\ref{app:check}.

\begin{figure}[t]
    \centering
    \includegraphics[width=\linewidth]{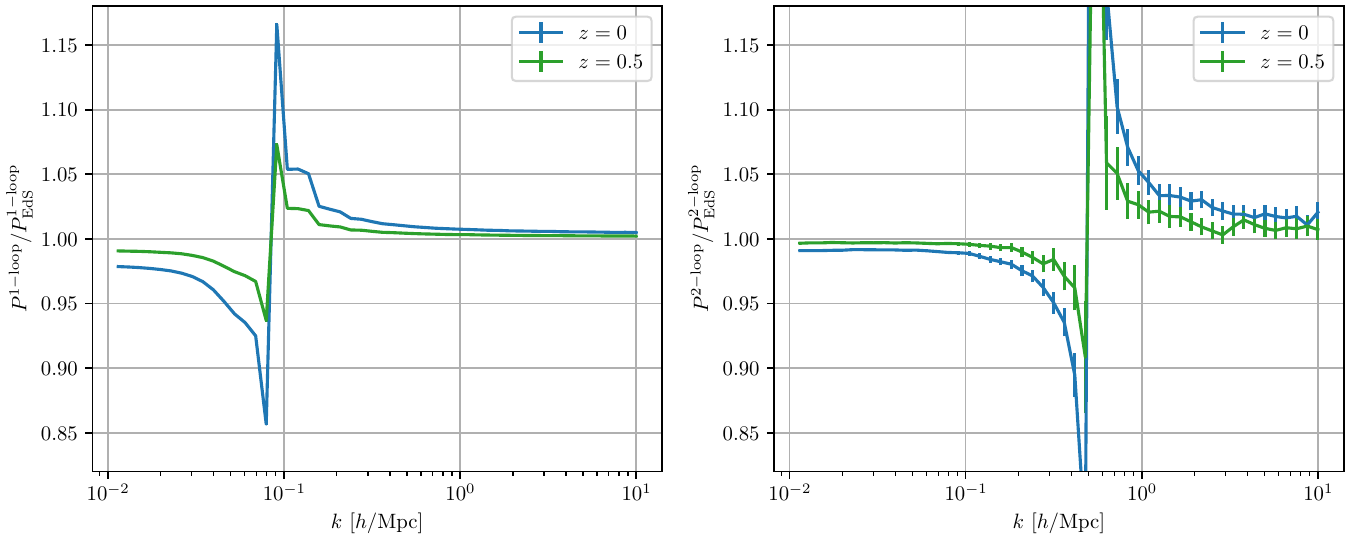}
    \caption{Non-linear corrections to the matter power spectrum with exact
        time-dependence, normalized to the corresponding terms in the EdS
        approximation. \emph{Left:} 1-loop correction and \emph{right:} 2-loop
        correction. The blue curves correspond to $z = 0$ while the green ones
        correspond to $z = 0.5$. The error bars display the uncertainty from
        the numerical integration. Note that the spikes at $k \simeq
        0.08~h/\Mpc$ (1-loop) and $k \simeq 0.5~h/\Mpc$ (2-loop) are due to the
        non-linear corrections crossing zero, leading to large relative
        deviations.}
    \label{fig:lcdm_vs_eds_L12}
\end{figure}

\begin{figure}[t]
    \centering
    \includegraphics[width=\linewidth]{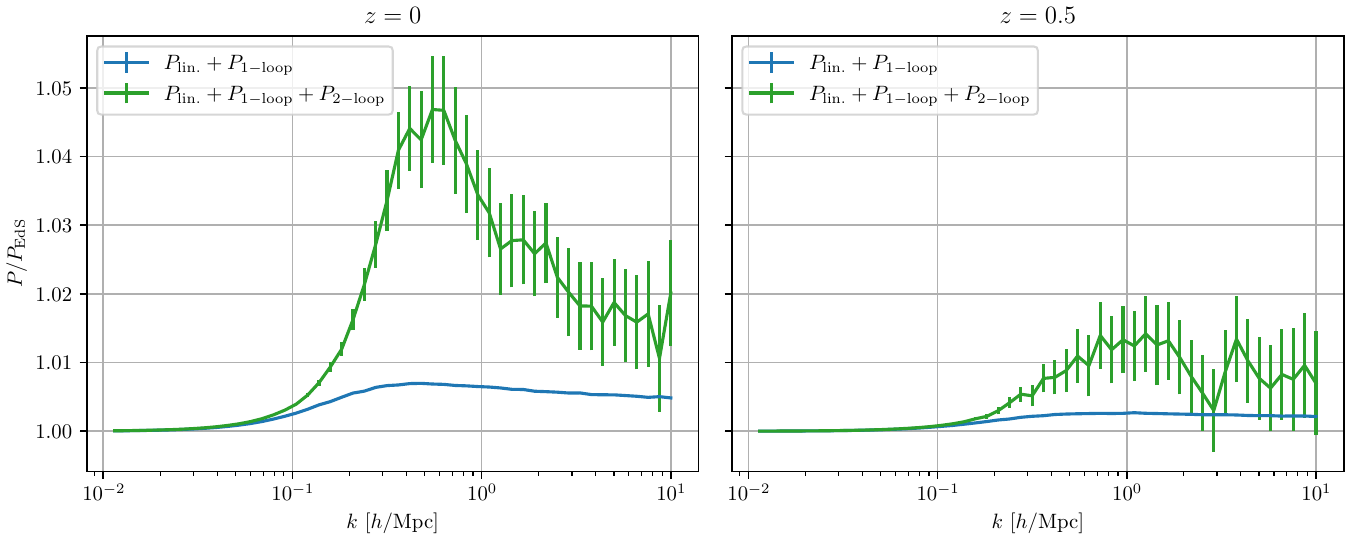}
    \caption{Total matter power spectrum at 1- and 2-loop with exact
    time-dependence normalized to that in the EdS approximation. \emph{Left:}
Redshift $z = 0$. \emph{right:} $z=0.5$. The uncertainty from the numerical
integration is indicated with error bars.}
    \label{fig:lcdm_vs_eds_tot}
\end{figure}

We compute the 1- and 2-loop corrections to the power spectrum both in the EdS
approximation and with the exact time-dependence and compare the results. In
Fig.~\ref{fig:lcdm_vs_eds_L12}, this comparison is shown for each loop
correction separately and at redshifts $z = 0$ and $z = 0.5$, respectively. At
$z = 0.5$ the EdS approximation works better, since the exact solution only
very recently starts to deviate from the EdS one. At $k \simeq 0.08~h/\Mpc$,
the 1-loop correction switches sign and crosses zero, which yield large
relative deviations (similarly for the 2-loop term at $k \simeq 0.5~h/\Mpc$).
Fig.~\ref{fig:lcdm_vs_eds_tot} displays the total power spectrum at 1- and
2-loop normalized to the EdS solution, computed at two redshifts $z = 0$ and
$z = 0.5$. On large scales the linear contribution dominates, which is not
affected by the EdS approximation, hence the departure from EdS starts to
appear around $k \simeq 0.1~h/\Mpc$. In addition to the EdS approximation being
more reliable at earlier times, the non-linear corrections are smaller compared
to the linear power spectrum at $z = 0.5$, hence the deviation of the total
power spectrum is only a few permille in the weakly non-linear regime of the
right plot.

\section{Two-fluid model: CDM + baryons and massive neutrinos}
\label{sec:2fluid_lin}

In this section we  set up a two-component fluid model for baryons, CDM and
massive neutrinos, and discuss its validity. Baryons and CDM are treated
equally as a single component, forming a perfect, pressureless fluid. They
interact through gravity with the neutrinos, which are described by a
fluid with an effective sound velocity. A fluid model for neutrinos
was thoroughly studied in Ref.~\cite{Shoji:2010}, which found that the fluid
description was only accurate at the level of 10--20\%. We aim for far better
accuracy than that, hence we follow Ref.~\cite{Blas:2014} and use a hybrid
scheme: at high redshift we use the full Boltzmann hierarchy while at low
redshift we switch to the two-fluid model. The reasoning behind this is
two-fold. Firstly, after the neutrinos become non-relativistic, the coupling
between lower and higher moments of the Boltzmann hierarchy is suppressed by
powers of $T_{\nu} / \mnu$. This decoupling allows us to follow only the lowest
moments at late times. Secondly, non-linear corrections become important at
late redshift $z \lesssim 10$, and computing them numerically is manageable in
the two-fluid model. Therefore, we match the Boltzmann solution onto the
two-fluid model at $\zmatch = 25$, which for the neutrino masses we
consider is well after the non-relativistic transition given by
\begin{equation}
    z_{\nr} \simeq 1890 \, \frac{\mnu}{1~\eV}\, .
\end{equation}
We find that depending on the specific expression used for the neutrino
effective sound velocity, the two-fluid model agrees with the full Boltzmann
solution at (sub-)permille and percent level for the CDM+baryons and neutrino
transfer functions, respectively (see below).

We consider four models with neutrino masses $\sum \mnu = 0.06$, $0.15$, $0.21$
and $0.3~\eV$, respectively. In all models we have three degenerate neutrino
species, described jointly by a single fluid component. In principle it is
possible to extend the formalism to a multi-fluid where each neutrino species
has an individual effective sound velocity. However, cosmological probes are quite
insensitive to the specific details of neutrino physics except for the absolute
mass scale, and the error introduced by using degenerate neutrino species
appears to be negligible compared to expected sensitivity of future surveys,
see e.g.~Ref.~\cite{Archidiacono:2020dvx}.

Neglecting vorticity and dropping non-linear terms for the moment, the
continuity and Euler equations for the CDM+baryons and neutrino density contrast
and velocity divergence are
\begin{subequations}
    \begin{align}
        \partial_{\tau} \delta_{\cb} + \theta_{\cb} & = 0\, , \\
        \partial_{\tau} \theta_{\cb} + \mH \theta_{\cb}
            + \frac{3}{2} \mH^2 \Omega_m
            [\fnu \delta_{\nu} + (1 - \fnu) \delta_{\cb}] & = 0\, , \\
        \partial_{\tau} \delta_{\nu} + \theta_{\nu} & = 0\, , \\
        \partial_{\tau} \theta_{\nu} + \mH \theta_{\nu}
            + \frac{3}{2} \mH^2 \Omega_m
            [\fnu \delta_{\nu} + (1 - \fnu) \delta_{\cb}]
        - k^2 c_s^2 \delta_{\nu} + k^2 \sigma & = 0\, .
        \label{eq:2fluid_cont_euler:nu_euler}
    \end{align}
    \label{eq:2fluid_cont_euler}
\end{subequations}
Here, the combined baryon and CDM density contrast is
\begin{equation*}
    \delta_{\cb} = \frac{ f_{\mathrm{b}}\delta_{\mathrm{b}} + f_{\cdm} \delta_{\cdm}}
        {f_{\mathrm{b}} + f_{\cdm}}\,,
\end{equation*}
where the energy density fraction for each species is $f_i \equiv \Omega_i /
\Omega_m$. The velocity divergence $\theta_{\cb}$ is defined equivalently. The
energy density fraction of neutrinos is constant for $z \ll z_{\nr}$,
\begin{equation}
    \fnu = \frac{1}{\Omega_m(z=0)} \frac{\sum \mnu}{93.14~h^2\eV}\, .
\end{equation}
Specifically, this gives $\fnu = 0.0045$, $0.0112$, $0.0156$, and $0.0221$ for
$\sum \mnu = 0.06$, $0.15$, $0.21$ and $0.3~\eV$, respectively. For the lowest
neutrino masses, the exact neutrino fraction $\fnu(z)$ differs about 1\% around
$\zmatch = 25$ from the constant approximation above, however taking the exact
time-dependence into account only alters the transfer functions at $z = 0$ at
the order of $10^{-4}$, hence we neglect it.

Lastly, in the Euler equation for the neutrino component,
Eq.~\eqref{eq:2fluid_cont_euler:nu_euler}, the sound velocity term expresses the
free-streaming of the neutrinos:
$c_s^2 = \delta P_{\nu}/ \delta\rho_{\nu}$. In addition, $\sigma$ denotes the
anisotropic stress. Both $c_s^2$ and $\sigma$ needs to be specified in order to
close the system in Eqs.~\eqref{eq:2fluid_cont_euler}, and we return to this issue
below.

Again, we find it convenient to rewrite the continuity and Euler equations in
terms of $\etaD = \ln D(z)$. However, for a model with massive neutrinos the
growth factor is $k$-dependent, which complicates the analysis. To avoid this
issue, we instead use the growth factor $D_{\fnu=0}(z)$ of a corresponding
reference cosmology with massless neutrinos. Specifically, given a cosmology
with massive neutrinos, the reference cosmology has the exact same parameters
except for the CDM energy density which is adjusted such that the total matter
energy densities of the models are the same. The equation for the growth factor
$D_{\fnu=0}(z)$ is
\begin{equation}
    \frac{\dd^2 D_{\fnu = 0}(z)}{\dd \ln a^2}
    +
    \left(
        1 + \frac{\dd \ln \mH}{\dd \ln a}
    \right)
    \frac{\dd D_{\fnu = 0}(z)}{\dd \ln a}
    - \frac{3}{2} \Omega_m D_{\fnu = 0}(z)
    = 0\, .
\end{equation}
For scales much larger than the neutrino free-streaming scale,
neutrinos behave as CDM and the growth factors of the two
models agree. We define the tuple $\psi_a$ as
\begin{equation}
    \psi_1 = \delta_{\cb} \, , \quad
    \psi_2 = - \frac{\theta_{\cb}}{\mH f} \, , \quad
    \psi_3 = \delta_{\nu} \, , \quad
    \psi_4 = - \frac{\theta_{\nu}}{\mH f} \, ,
\end{equation}
where $f = \dd \ln D_{\fnu = 0} /\dd \ln a$. Using
$\partial_{\tau} = \mH f \, \partial_{\etaD}$, the linearized continuity and Euler
equations become
\begin{equation}
    \partial_{\etaD} \psi_a + \Omega(k,\etaD)_{ab} \, \psi_b = 0\,,
    \label{eq:2fluid_lin_compact}
\end{equation}
with
\begin{equation}
    \Omega(k, \etaD) =
    \begin{pmatrix}
        0 & -1 & 0 & 0 \\
        - \frac{3}{2} \omegaoverfsq (1 - \fnu) &
        \frac{3}{2} \omegaoverfsq - 1 &
        - \frac{3}{2} \omegaoverfsq \fnu & 0 \\
        0 & 0 & 0 & -1 \\
        - \frac{3}{2} \omegaoverfsq (1 - \fnu) &
        0 &
        - \frac{3}{2} \omegaoverfsq [\fnu - k^2 c_{s,\mathrm{eff}}^2(k,\etaD) ] &
        \frac{3}{2} \omegaoverfsq - 1
    \end{pmatrix}
    \, ,
    \label{eq:2fluid_omega_mat}
\end{equation}
where $c_{s,\mathrm{eff}}^2$ is an effective sound velocity that will be discussed
shortly. In solving Eq.~\eqref{eq:2fluid_lin_compact}, we write
$\psi_a(\vk,\etaD) = F_a^{(1)}(k,\etaD)\, \delta_{\cb}(\vk,\etaDmatch)$ and use the
growing mode eigenvector of the matrix $\Omega(k,\etaDmatch)$ as initial
condition for the linear kernels $F_a^{(1)}$ (suitably normalized so that
$F_1^{(1)}(k,\etaDmatch) = 1$). We will clarify the reasoning for this choice
of initial conditions when we move on to non-linear corrections in the next
section. Thus, we may compute e.g.\ the linear CDM+baryons and neutrino cross
power spectrum at some time $\etaD$ by
\begin{equation}
    P_{\cb,\nu}(k,\etaD) = F_1^{(1)}(k,\etaD)\, F_3^{(1)}(k,\etaD)
        \, P_{\cb,\cb}(k, \etaDmatch)\, ,
\end{equation}
where $P_{\cb,\cb}(k,\etaDmatch)$ is taken from CLASS.

\subsection{Sound velocity and anisotropic stress}
We will now describe two ways to close the system in
Eq.~\eqref{eq:2fluid_lin_compact}: (i) by approximating the sound velocity by the
adiabatic sound velocity and neglecting the anisotropic stress and (ii) by
integrating the Boltzmann equation for neutrinos numerically to obtain the
exact sound velocity and anisotropic stress. Both methods will be used and
compared when we discuss non-linear corrections in the next section.

\paragraph{Adiabatic approximation (2F-ad)}

In the non-relativistic limit the sound velocity of the neutrinos approaches the
adiabatic sound velocity $c_g$, which can be related to the velocity dispersion
$\sigma_{\nu}$ (assuming $\sigma_{\nu}^2 \ll 1$)~\cite{Shoji:2010}:
\begin{equation}
    c_s^2 \equiv \frac{\delta P}{\delta \rho} \xrightarrow[z \ll z_{\nr}]{}
    c_g^2 = \frac{5}{9} \sigma_{\nu}^2 =
    \frac{25}{3} \frac{\zeta (5)}{\zeta (3)}
    \left(
        \frac{T_{\nu}}{\mnu}
    \right)^2 \, .
\end{equation}
Assuming also that the anisotropic stress can be neglected in this limit, the
effective sound velocity entering the two-fluid equations in
Eqs.~\eqref{eq:2fluid_lin_compact} and \eqref{eq:2fluid_omega_mat} is
\begin{equation}
    c_{s,\mathrm{eff}}^2 = \frac{2}{3 \Omega_m \mH^2} c_g^2
    \simeq \frac{1.214}{\Omega_m(z=0)}
    \left(
        \frac{1~\eV}{\mnu}
    \right)^2
    (1+z)\, \frac{\Mpc^2}{h^2}\, .
\end{equation}
We will refer to this scheme in the future with the label 2F-ad.
It has the benefit of being simpler to implement in practice compared
to the more accurate scheme that we turn to next.

\paragraph{Exact effective sound velocity (2F)}

The idea of this scheme is to obtain an optimal fluid description that is
informed about the impact of the linear perturbations to the complete neutrino
distribution function, including all of its higher moments. This is realized by
using the exact neutrino sound velocity and anisotropic stress in the Euler
equation, derived from a full solution of the linearized collision-less
Boltzmann equation~\cite{Ma:1995ey}
\begin{equation}
    \left(
        \partial_{\tau} + \frac{\im \, q}{\epsilon(q, \tau)} (\vk\cdot\vnhat)
    \right)
    \Psi(\vk,\vnhat, q, \tau)
    + \frac{\dd \ln f_0}{\dd \ln q}
    \left(
        \partial_{\tau} \phi(k,\tau)
        - \frac{\im \, \epsilon(q,\tau)}{q} (\vk\cdot\vnhat) \psi(k,\tau)
    \right)
    = 0\, .
    \label{eq:boltzmann}
\end{equation}
Here, $\Psi$ is the linear perturbation to the neutrino distribution function,
$f = f_0 (1 + \Psi)$, $\vnhat = \vq/q$ and $\phi$ and $\psi$ are metric
perturbations in the conformal Newtonian gauge. Furthermore, $q$ and
$\epsilon(q,\tau) = \sqrt{q^2 + a^2(\tau) m^2}$ are the comoving momentum
and energy of a particle, respectively. Neutrinos decoupled while they
were ultra-relativistic, hence the unperturbed distribution function $f_0$
remains in the form of an ultra-relativistic Fermi-Dirac distribution also
after the non-relativistic transition. Integrating Eq.~\eqref{eq:boltzmann} and
projecting onto Legendre polynomials $P_l(\vk\cdot\vnhat/k)$ yields the
following multipole solutions~\cite{Shoji:2010}:
\begin{multline}
    \Psi_l (k,q,\tau) =
    \frac{\dd \ln f_0(q)}{\dd \ln q}
    \times \Bigg(
    \phi(k, \tau_i) \, j_l[k\, y(\tau_i,\tau)] - \phi(k, \tau) \delta_{l0}
    \\
    - k \int_{\tau_i}^{\tau} \dd \tilde{\tau} \,
    \left[
        \frac{\epsilon(q,\tilde{\tau})}{q} \psi(k, \tilde{\tau})
        + \frac{q}{\epsilon(q,\tilde{\tau})} \phi(k, \tilde{\tau})
    \right]
    \left[
        \frac{l}{2l + 1} \, j_{l-1} [k\, y(\tilde{\tau}, \tau)]
        - \frac{l+1}{2l + 1} \, j_{l+1} [k\, y(\tilde{\tau},\tau)]
    \right]
    \Bigg)
    \\
    + \sum_{l'}\sum_{l''} (-\im)^{l' + l'' - l} (2l' + 1) (2l'' + 1) \,
    \Psi_{l'}(k, q, \tau_i) \, j_{l''}[k\, y(\tau_i, \tau)]
    \begin{pmatrix}
        l & l' & l'' \\
        0 & 0 & 0
    \end{pmatrix}^2
    \, ,
    \label{eq:psi_analytic_sol}
\end{multline}
where $j_l$ are spherical Bessel functions,
$
\begin{pmatrix}
        l & l' & l'' \\
        0 & 0 & 0
\end{pmatrix}
$
is the Wigner 3-j symbol, and
\begin{equation}
    y(\tau_a, \tau_b) = \int_{\tau_a}^{\tau_b} \dd \tau \,
        \frac{q}{\epsilon(q,\tau)}\, .
\end{equation}
Note that we fixed a sign typo from Ref.~\cite{Shoji:2010} in the $\phi$-term of the second line of Eq.~\eqref{eq:psi_analytic_sol}.

The perturbations $\Psi_l$ are related to the density contrast, velocity
divergence and anisotropic stress through integrals over momentum:
\begin{subequations}
    \begin{align}
        \bar{\rho}(\tau) & = \frac{4\pi}{a^4(\tau)} \int \dd q \, q^2 \epsilon(q, \tau) f_0(q)\, , \\
        \bar{P}(\tau) & = \frac{4\pi}{3 a^4(\tau)} \int \dd q \, q^2 \frac{q^2}{\epsilon (q, \tau)} f_0(q)\, , \\
    \delta\rho(k, \tau) & = \frac{4\pi}{a^4(\tau)} \int \dd q \, q^2 \epsilon(q, \tau) f_0(q) \Psi_0(k,q,\tau)\, , \\
    \delta P(k, \tau) & = \frac{4\pi}{3 a^4(\tau)} \int \dd q \, q^2 \frac{q^2}{\epsilon (q, \tau)} f_0(q) \Psi_0(k, q, \tau)\, , \\
    (\bar{\rho} + \bar{P}) \theta(k,\tau) & = \frac{4\pi k}{a^4(\tau)} \int \dd q \, q^3 f_0(q) \Psi_1(k,q,\tau)\, , \\
    (\bar{\rho} + \bar{P}) \sigma(k,\tau) & = \frac{8\pi}{3 a^4(\tau)} \int \dd q \, q^2 \frac{q^2}{\epsilon (q, \tau)} f_0(q) \Psi_2(k,q,\tau) \, .
    \end{align}
    \label{eq:fluid_quantities}
\end{subequations}
Thus, using Eqs.~\eqref{eq:psi_analytic_sol} and \eqref{eq:fluid_quantities} we
may compute the exact sound velocity $c_s^2(k,\tau) \equiv \delta P/\delta\rho$,
density perturbation $\delta(k,\tau) \equiv \delta\rho/\bar{\rho}$ and anisotropic
stress $\sigma(k,\tau)$ to obtain the effective sound velocity which is used in the
two-fluid evolution,
\begin{equation}
    c_{s,\mathrm{eff}}^2 = \frac{2}{3 \Omega_m \mH^2}
    \left(
        c_s^2(k,\tau) - \frac{\sigma(k,\tau)}{\delta(k,\tau)}
    \right) \, .
    \label{eq:eff_cs2}
\end{equation}

In practice, we simplify Eq.~\eqref{eq:psi_analytic_sol} by assuming $\phi =
\psi$ at all times and $\psi = \mathrm{const}$ before vacuum energy domination
(specifically, before $z = 5$). In addition, we neglect the contributions from
the initial perturbations $\Psi_l(k,q,\tau_i)$ since the integral term
dominates at late times. The gravitational potential is obtained from a
Boltzmann solver (CLASS) and $\tau_i = 1~\Mpc$ is used. We check explicitly
for a subset of wavenumbers and the neutrino masses we consider that dropping
these assumptions has negligible impact on the numerical results.

The sound velocity and anisotropic stress may also be extracted from
CLASS, which solves the Boltzmann equation \eqref{eq:boltzmann} by
immediately projecting onto Legendre polynomials and integrating the resulting
hierarchy of differential equations for the $\Psi_l$'s.%
\footnote{In our utilization of CLASS for cosmologies with massive neutrinos,
we turn off the default fluid approximation for neutrinos, i.e.\ we use
\texttt{ncdm\_fluid\_approximation = 3} and \texttt{l\_max = 17} unless stated
otherwise.}
A truncation of the hierarchy at some maximum multipole $l_{\mathrm{max}}$ is
needed in order to be able to solve it numerically, which may introduce
inaccuracies that propagate the hierarchy to lower multipoles. Note that this
problem is not present in Eq.~\eqref{eq:psi_analytic_sol}, since there is no
coupling between multipoles (up to backreaction on the gravitational
potentials). Nevertheless, using Eq.~\eqref{eq:psi_analytic_sol} for the
neutrino perturbation in a Boltzmann solver would be very time-consuming, since
the integral has to be performed at every time step in order to update transfer
functions and metric perturbations (we avoid this by taking the metric
perturbations as external input).

\begin{figure}[t]
    \centering
    \includegraphics[width=0.49\linewidth]{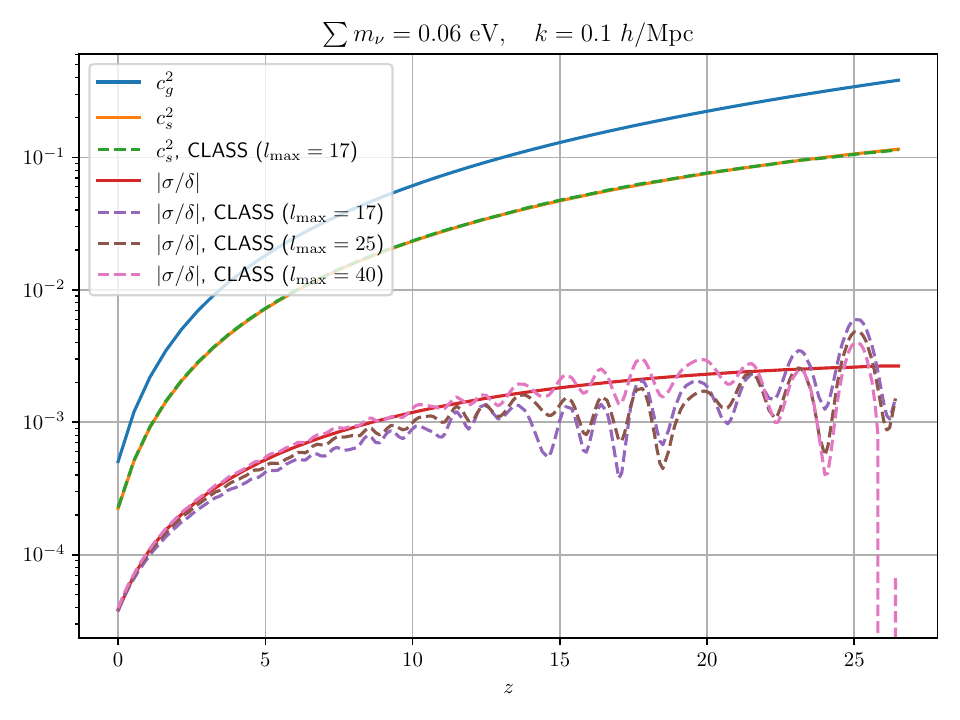}
    \includegraphics[width=0.49\linewidth]{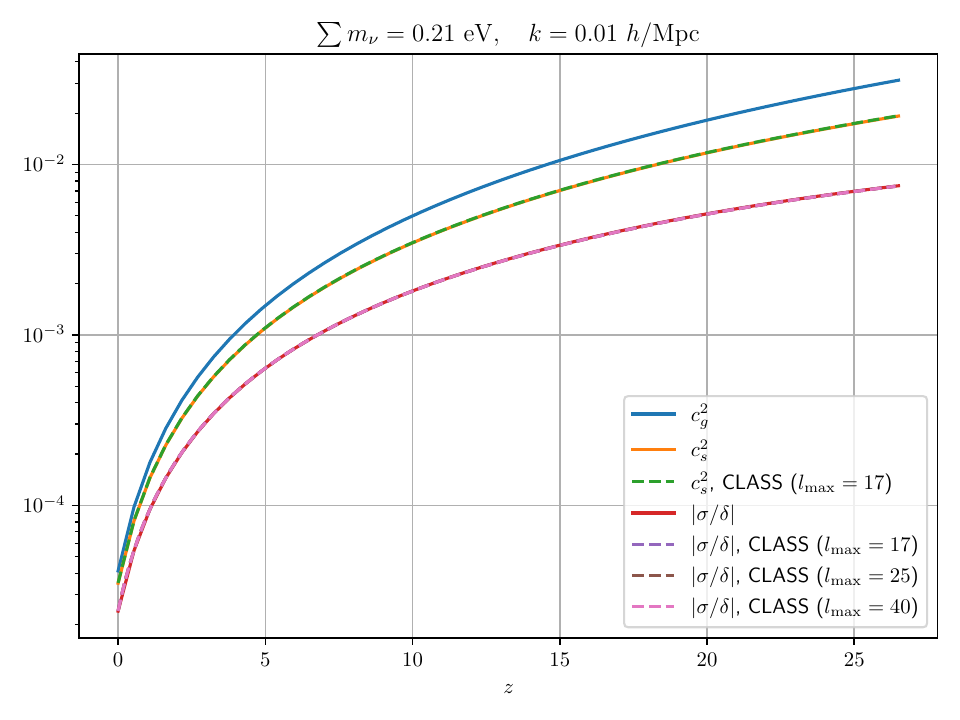}
    \caption{Adiabatic sound velocity $c_g^2$, sound velocity $c_s^2$ and absolute
        value of the anisotropic stress divided by the density contrast,
        $|\sigma/\delta|$, as a function of redshift from our numerical
        computation and from CLASS using different maximum number of multipoles
        $l_{\mathrm{max}}$ (note that the difference is invisible except for
        $|\sigma/\delta|$ in the left plot). \emph{Left:} neutrino mass
        $\sum\mnu = 0.06~\eV$ and wavenumber $k = 0.1~h/\Mpc$.
    \emph{Right:} $\sum\mnu = 0.21~\eV$ and $k = 0.01~h/\Mpc$.}
    \label{fig:cg2_cs2_sigma_over_delta}
\end{figure}

In Fig.~\ref{fig:cg2_cs2_sigma_over_delta}, we compare our numerical results
for the sound velocity $c_s^2$ and anisotropic stress over density contrast
$|\sigma/\delta|$ with CLASS for two choices of neutrino mass and
wavenumber. In addition, the adiabatic sound velocity $c_g^2$ is plotted. We see
that the sound velocity is in good agreement with CLASS; a closer quantitative
analysis yields agreement of around 1\%, also for a larger set of wavenumbers
and neutrino masses, and independent of the maximum number of multipoles
$l_{\mathrm{max}}$ used in CLASS. Furthermore, for high neutrino masses and low
wavenumbers, the ratio of anisotropic stress and density contrast also has
percent level agreement with that from CLASS. In the other limit, the CLASS
solution develops oscillatory behavior, and only the mean value of the
oscillations follows our numerical result. This is seen for $k = 0.1~h/\Mpc$ in
the left plot of Fig.~\ref{fig:cg2_cs2_sigma_over_delta}. For even smaller
scales, the oscillations become more pronounced and results using different
multipole truncation number $l_\mathrm{max}$ do not converge to the same
answer even at late times. Nonetheless, in this limit $|\sigma/\delta|$ is
suppressed compared to the sound velocity, so the impact on the velocity
divergence in the Euler equation is small. In principle, we could have used
CLASS to compute the sound velocity and anisotropic stress entering
Eq.~\eqref{eq:eff_cs2} and in turn entering the two-fluid model, but the
oscillatory behavior on small scales makes this impractical, hence we opt for
numerically solving Eqs.~\eqref{eq:psi_analytic_sol} and
\eqref{eq:fluid_quantities}.

In the following we will use the label 2F for the two-fluid model with the
exact effective sound velocity in Eq.~\eqref{eq:eff_cs2}.

\subsection{Comparison with Boltzmann solver}

\begin{figure}[t]
    \centering
    \includegraphics[width=0.6\linewidth]{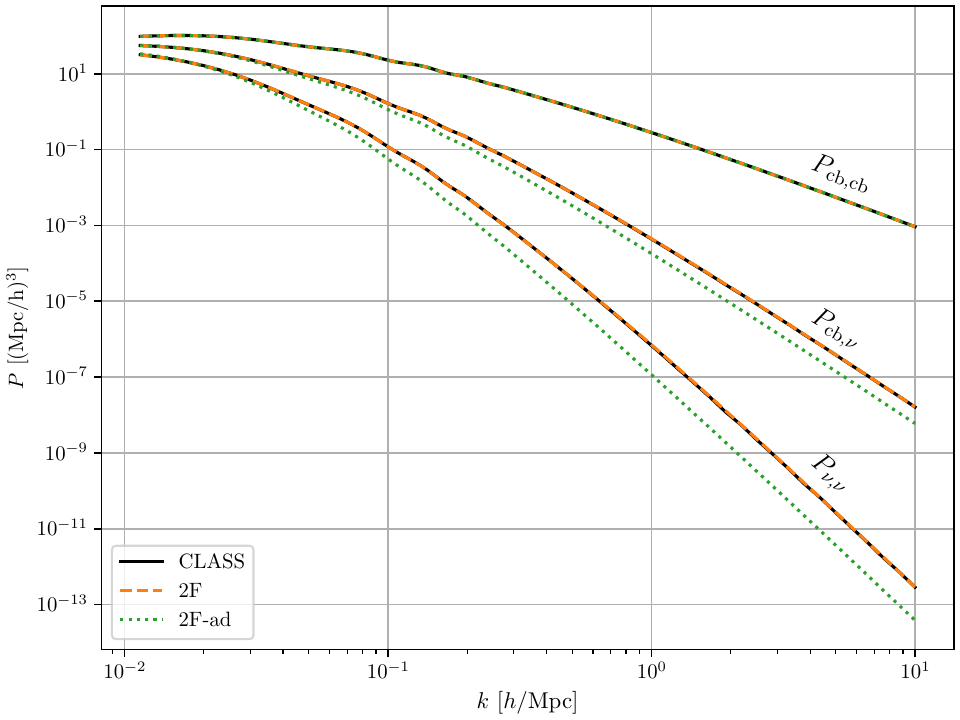}
    \caption{Contributions to the linear matter power spectrum at $z = 0$ in a
    cosmology with $\sum\mnu = 0.15~\eV$. The solid black lines correspond to
the CLASS solution, the dashed orange lines to the exact two-fluid solution and
the dotted green lines to the two-fluid solution using the adiabatic
approximation.}
    \label{fig:linear_all_m_nu_005}
\end{figure}

Having set up the two-fluid model for CDM+baryons and massive neutrinos in the
previous subsection, we proceed to compare this scheme at the linear level
with a Boltzmann solver (CLASS). The total matter power spectrum is the
weighted sum of the CDM+baryons power spectrum, the neutrino power spectrum and
the cross power spectrum between CDM+baryons and neutrinos:
\begin{equation}
    P_{\mm} = (1 - \fnu)^2 P_{\cb,\cb} + 2 (1 - \fnu) \fnu P_{\cb,\nu}
        + \fnu^2 P_{\nu,\nu}\, .
    \label{eq:PS_mm}
\end{equation}
The various contributions are shown in Fig.~\ref{fig:linear_all_m_nu_005} at
redshift $z = 0$ and for neutrino mass $\sum\mnu = 0.15~\eV$, where we include
the results from the two-fluid model (both using adiabatic approximation and
exact sound velocity) and from CLASS. The neutrino power spectrum and the cross
power spectrum are suppressed compared to the CDM+baryons power
spectrum due to free-streaming of the neutrinos on scales smaller than the
free-streaming scale
$k_{\mathrm{FS}} = 1/c_{s,\mathrm{eff}}^2 \sim 10^{-2}~h/\Mpc$. Since in
addition $\fnu \ll 1$, the total matter power spectrum is dominated by the
$P_{\cb,\cb}$ term on these scales. However, through the backreaction on the
gravitational potential, the presence of massive, free-streaming neutrinos
leads to a reduction of growth in the baryon and CDM sector, which at the linear
level yields the well known $\sim 8 \fnu$ reduction of the total power
spectrum. Therefore it is necessary to calculate the neutrino transfer function
accurately even when neglecting the last terms of Eq.~\eqref{eq:PS_mm}.

\begin{figure}[t]
    \centering
    \includegraphics[width=\linewidth]{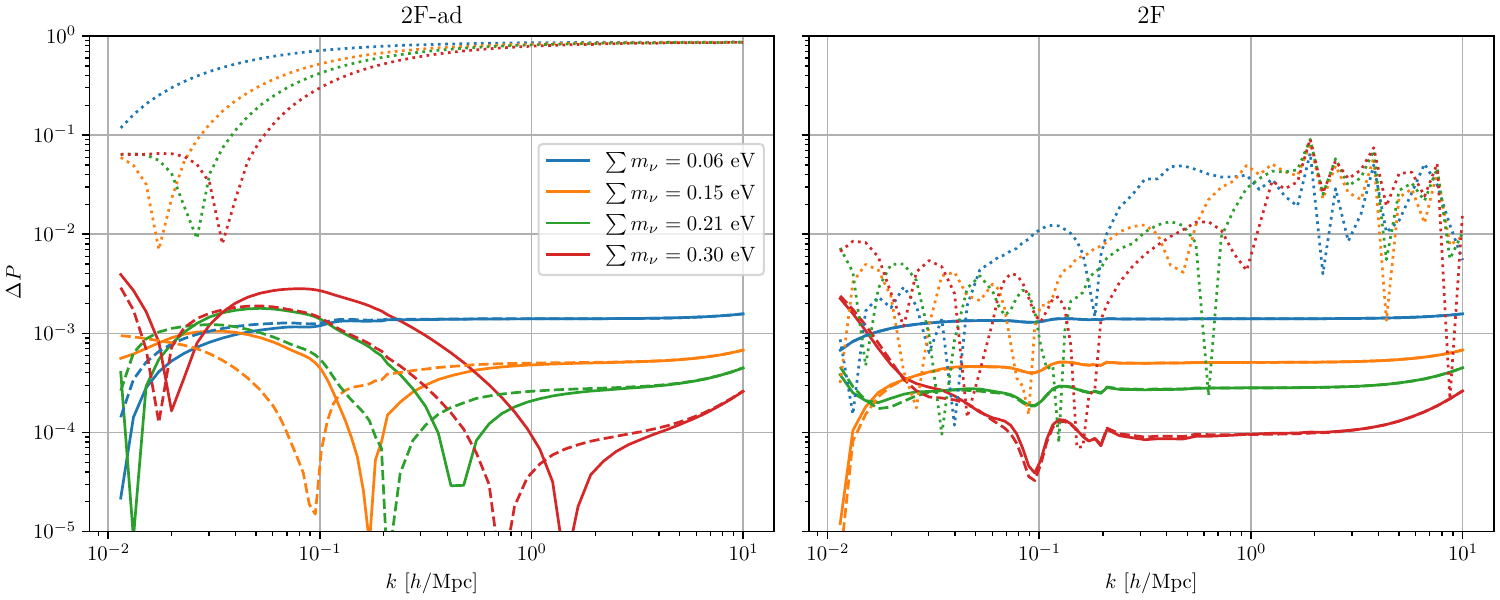}
    \caption{Relative difference between the linear power spectrum computed in
        the two-fluid model and the Boltzmann solution,
        $\Delta P = |P^{\mathrm{Two-fluid}}/P^{\mathrm{Boltzmann}} - 1|$,
        for various neutrino masses at $z = 0$. The solid lines correspond to
        the total matter power spectrum $\Delta P_{\mm}$, the dashed lines
        correspond to the CDM+baryons power spectrum $\Delta P_{\cb,\cb}$ and the
        dotted lines correspond to the neutrino power spectrum $\Delta
        P_{\nu,\nu}$. \emph{Left:} Two-fluid model using the adiabatic sound
        velocity approximation (scheme 2F-ad). \emph{Right:} Two-fluid model
    using exact sound velocity and anisotropic stress (scheme 2F).}
    \label{fig:linear_vs_CLASS}
\end{figure}

Fig.~\ref{fig:linear_vs_CLASS} displays the relative difference between the
two-fluid model and the CLASS solution for the neutrino masses we consider
(\emph{left:} adiabatic approximation, \emph{right:} exact sound velocity). We see
that the two-fluid model with adiabatic approximation underestimates the
neutrino-neutrino and cross power spectra by a significant amount (this is
also evident from Fig.~\ref{fig:linear_all_m_nu_005}). The decrease in power is
caused by the adiabatic approximation overestimating the sound velocity, as is
apparent from Fig.~\ref{fig:cg2_cs2_sigma_over_delta}, leading to suppressed
growth on scales smaller than the free-streaming scale. On scales close to and
smaller than the free-streaming scale $k_{\mathrm{FS}}$, the Poisson term
begins to dominate in the neutrino Euler equation, so the performance of the
adiabatic model improves. We note that the adiabatic approximation works better
for increasing neutrino masses, where a larger fraction of the neutrino
distribution has become sufficiently non-relativistic, and the sound velocity is
further in approaching the adiabatic limit. On the other hand, when $\fnu$ is
larger the error in the neutrino sector affects the total matter power spectrum
to greater extent: the gravitational coupling between CDM+baryons and neutrinos
is larger and the last two terms in Eq.~\eqref{eq:PS_mm} are more important.
Nevertheless, for the total power spectrum and the neutrino masses we consider,
we find agreement between the adiabatic model and the CLASS solution of a few
permille.

As expected, the two-fluid model with exact effective sound velocity agrees
excellently with the Boltzmann solution, with deviations of the order of
$10^{-4}$--$10^{-3}$ depending on the neutrino mass, as seen in the right plot of
Fig.~\ref{fig:linear_vs_CLASS}. The neutrino power spectrum agrees with the
Boltzmann solution from CLASS at percent level or better for $k\sim 0.1~h/\Mpc$,
and at the few percent level for $k\gtrsim 1~h/\Mpc$. However in the latter region the
Boltzmann solution from CLASS also differs by a few percent depending on the
maximum number of multipoles $l_{\mathrm{max}}$ included in the Boltzmann
hierarchy, indicating that inaccuracies due to the truncation of the hierarchy
propagate to lower multipoles.

Finally, we comment on a few alternative approximations
for neutrinos that truncate the Boltzmann hierarchy at $l_{\mathrm{max}} = 2$,
and hence include an evolution equation for the anisotropic stress in the fluid
equations. Various trunction schemes have been used for $\Psi_3$ to close the
system: the simplest one is $\Psi_3 = 0$~\cite{Shoji:2010}, but this is rather
inaccurate and the errors propagate to lower multipoles. An improved version
is to use the spherical Bessel function recurrence relation~\cite{Ma:1995ey},
and a similar scheme is used by the default neutrino treatment in
CLASS~\cite{Lesgourgues:2011rh}. Nevertheless, the neutrino power spectrum is
reduced by a factor $\mathcal{O}(10\%)$ around $k = 0.1~h/\Mpc$ using this
truncation scheme, compared to using the full Boltzmann hierarchy. It was noted
in Ref.~\cite{Archidiacono:2015ota} that due to the gravitational source terms,
the Bessel relation poorly captures the behavior of $\Psi_3$, and using instead
an empirical formula yields percent agreement for the neutrino power spectrum.
Our two-fluid model goes even further in that it captures the effect of all
higher order cumulants effectively through the sound velocity and anisotropic
stress in the Euler equation. In this sense, the agreement with the full
Boltzmann hierarchy is largely a consistency check: the only difference between
the Boltzmann equations and the 2F scheme is the neglection of the neutrino
equation of state in the latter.

We conclude that the two-fluid model with neutrino sound velocity and
anisotropic stress computed from an exact solution of the linear Boltzmann
equation works at the $10^{-4}$ to $10^{-3}$ level for the linear total matter
power spectrum on scales $k\lesssim 1~h/\Mpc$, making it suitable for a
precision comparison at higher order in perturbation theory.

\section{Non-linear power spectrum with massive neutrinos}
\label{sec:2fluid_nonlin}

In this section we compute non-linear corrections to the matter power spectrum
in the presence of massive neutrinos. We use the two-fluid model described in
the previous section, and compare to other approximation schemes used in the
previous literature.

We use the algorithm presented in Sec.\,\ref{sec:extension} to compute the 1-
and 2-loop power spectrum with exact time- and scale-dependence, and specify
some details specific to massive neutrinos in the following.
Including non-linear terms in the two-fluid equations in
Eq.~\eqref{eq:2fluid_lin_compact} gives an equation of motion of the form
of Eq.\,\eqref{eq:eom_compact_general} where the non-zero vertices are
\begin{subequations}
    \begin{align}
        \gamma_{121}(\vk,\vk_1,\vk_2) =
        \gamma_{343}(\vk,\vk_1,\vk_2) \equiv \alpha(\vk_1,\vk_2)\,, \\
        \gamma_{222}(\vk,\vk_1,\vk_2) =
        \gamma_{444}(\vk,\vk_1,\vk_2) \equiv \beta(\vk_1,\vk_2)\,.
    \end{align}
\end{subequations}
We expand the density contrast and velocity divergence of the CDM+baryon
and neutrino components perturbatively according to Eq.\,\eqref{eq:Fan}.
Note that the linear evolution matrix $\Omega_{ab}(k,\etaD)$ is both time-
and scale-dependent due to the neutrino effective sound velocity.

To solve the differential equation \eqref{eq:eom_kernels}, we also need an
initial condition for the kernels at $\etaDmatch$. A simple initial condition
setup would be to extend what we did in the EdS analysis and use EdS-SPT and 0 as
initial conditions for the CDM+baryons and neutrino kernels, respectively. The
reasoning behind this would be that $\etaDmatch$ is deep in the matter dominated
era, where the EdS approximation is accurate, and the neutrino density contrast
and velocity divergence are much smaller than the CDM+baryons density contrast
in most of the integration domain, so that $F_3^{(n)}$,~$F_4^{(n)} \ll 1$.
However, this works poorly in practice since the initial condition excites
transient solutions that propagate the kernel hierarchy and do no entirely
decay away by $\etaD = 0$ ($z = 0$). In particular, the EdS linear growing mode
solution $(1,1)$ (which is also the linear growing mode solution of
Eq.~\eqref{eq:eom_compact}) is not the linear growing mode solution of
Eq.~\eqref{eq:2fluid_lin_compact}. Rather, the first two components of the
eigenvector of $\Omega(k,\etaD)$ in Eq.~\eqref{eq:2fluid_omega_mat}
corresponding to the growing mode differs by a factor $\mathcal{O}(0.01)$ to
$(1,1)$ depending on $\etaD$ and $k$.

Having this in mind, given the growing mode eigenvector $u_a(k)$ of
$\Omega(k,\etaDmatch)$, we set $F_a^{(1)}(k,\etaDmatch) = u_a(k)/u_1(k)$ as initial
condition for the linear kernels. Note that the normalization
implies $F_1^{(1)}(k,\etaDmatch) = 1$, consistent with the convention adopted
in Sec.~\ref{sec:2fluid_lin}, implying
$\delta_0(\vk)=\delta_{\cb}^{(1)}(\vk,\etaDmatch)$ in Eq.~\eqref{eq:Fan}.
Initializing the higher order kernels in the
growing mode is more tricky. In principle, it is possible to analytically solve
Eq.~\eqref{eq:eom_kernels} with $\Omega = \Omega(k,\etaDmatch)$
time-independent, obtaining a recursion solution similar to that in EdS-SPT
with a dependence on $k$. Instead, we let the system of differential equations
in the numerics reach the growing mode by extrapolating far back in time:
starting at $\etaDasymp = -10$ (corresponding to $z \sim 10^4$) we
set
\begin{equation}
    F_a^{(n)}(k,\etaDasymp) =
    \begin{cases}
        \frac{u_a(k)}{u_1(k)}\,  \e^{\lambda_1(\etaDasymp - \etaDmatch)}\, ,
            & n = 1 \,,\\
        0\, , & n > 1 \, ,
    \end{cases}
\end{equation}
where $\lambda_1$ is the growing mode eigenvalue of $\Omega(k,\etaDmatch)$,
and evolve the kernels to $\etaDmatch$ using $\Omega = \Omega(k,\etaDmatch)$.
This way, the linear kernels remain in the growing mode solution (by
construction) and any transient modes in the higher order kernels have decayed
by $\etaDmatch$. Thus, the entire kernel hierarchy resides in the growing mode
of $\Omega(k,\etaDmatch)$ when we turn on dynamics for $\etaD > \etaDmatch$.
Note that a similar strategy was used in the 1-loop analysis of \cite{Blas:2014}
to avoid transients that occur within the TRG framework \cite{Audren:2011ne}.

\begin{figure}[t]
    \centering
    \includegraphics[width=0.8\linewidth]{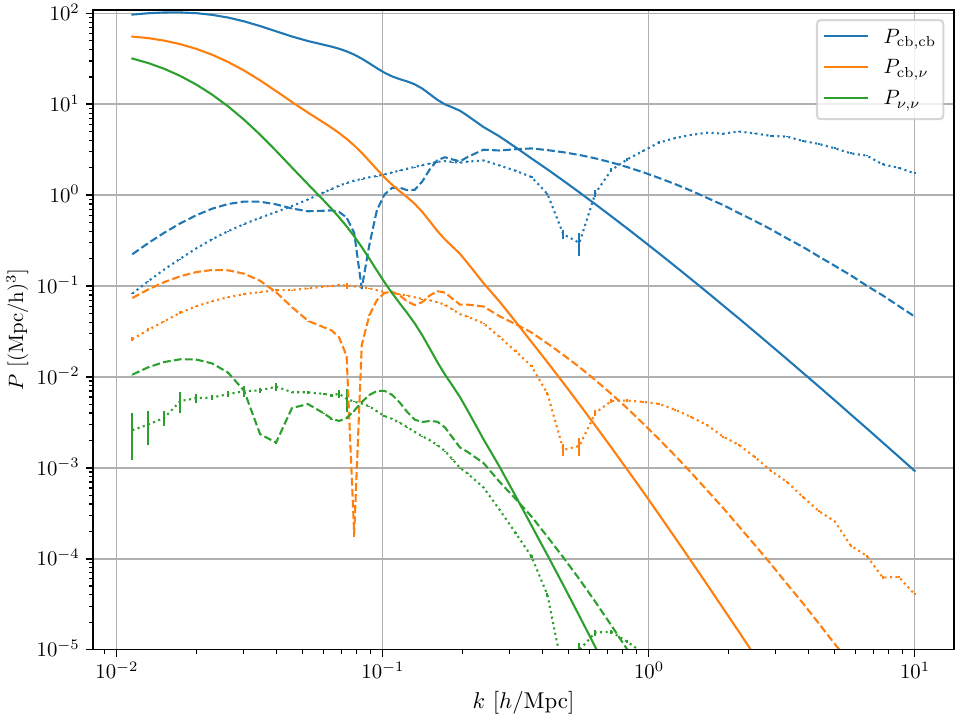}
    \caption{Contributions to the matter power spectrum at $z = 0$ computed in
        the 2F scheme for neutrino mass $\sum\mnu = 0.15~\eV$. Solid lines
        correspond to the linear contribution, while dashed and dotted lines
        correspond to the absolute value of the 1- and 2-loop corrections,
        respectively. Error bars (only visible in a few places) indicate uncertainty
    from the numerical integration.}
    \label{fig:abs_effcs2_m_nu005}
\end{figure}

The numerical analysis is based on a $\LCDM$ cosmology with massive neutrinos
and otherwise the same parameters as in Sec.~\ref{sec:eds_approx}:
$h = 0.6756$, $\Omega_{\mathrm{b}} = 0.04828$, $\Omega_{\cdm} = 0.2638$,
$n_s = 0.9619$ and $A_s = 2.215\cdot 10^{-9}$.
We take the input linear CDM+baryons power spectrum at $z_{\rm match}=25$
and the ratio $\Omega_m/f^2$ from CLASS. Fig.~\ref{fig:abs_effcs2_m_nu005}
shows the various contributions to the matter power spectrum at 2-loop with
neutrino mass $\sum\mnu = 0.15~\eV$, computed in the two-fluid model with exact
sound velocity (scheme 2F). As one would expect, the neutrino-neutrino and
CDM+baryons-neutrino cross correlation loop corrections are suppressed compared
to the corresponding CDM+baryons auto-correlation loop corrections. In addition,
the suppression is also present to a certain degree at scales larger than the
free-streaming scale, due to mode-mode coupling. Moreover, since they are
multiplied by $\fnu \ll 1$ (cf.~Eq.~\eqref{eq:PS_mm}), loop corrections to
$P_{\cb,\nu}$ and $P_{\nu,\nu}$ can safely be neglected for an analysis with
percent-level accuracy. Nevertheless, we include them for completeness in our
analysis.

\subsection{Approximation schemes}

We aim to identify viable simplified approximation schemes to capture the impact of
massive neutrinos on loop corrections within the weakly non-linear regime.
We therefore compare the following schemes, in the order of increasing complexity:%
\footnote{We do not consider the scheme of Ref.~\cite{Wong:2008ws}, due to
broken momentum conservation and consequently spurious behavior in the $k\to 0$
limit~\cite{Blas:2014}.}

\paragraph{EdS-SPT scheme (1F)}

The simplest scheme we consider takes neutrinos only into account via their
effect on the linear power spectrum,
\begin{equation}
    P_{\mm} = P_{\mm}^{\lin}
    + P_{\oneloop}^{\text{EdS-SPT}}(P_{\mm}^{\lin})
    + P_{\twoloop}^{\text{EdS-SPT}}(P_{\mm}^{\lin}) \, ,
\end{equation}
where $P_{\oneloop}^{\text{EdS-SPT}}(P_{\mm})$ symbolizes the 1-loop correction
computed using EdS-SPT kernels with $P_{\mm}$ as input; similarly for 2-loop.
While being particularly simple to implement, this scheme is completely
agnostic to the scale- and time-dependent free-streaming dynamics of the
neutrinos beyond the linear level. In addition, inaccuracies emerge at late
times due to the EdS approximation becoming increasingly unreliable (see
Sec.~\ref{sec:eds_approx}). We refer to this scheme as 1F in the following.

It turns out that the 1F-scheme underestimates the loop correction to the power
spectrum by a factor $\sim 2 L \fnu$, $L$ being the loop order, for scales much
smaller than the free-streaming scale. This will be discussed more below; at
this point we note that part of the motivation for the next scheme is to
correct for this.

\paragraph{EdS-SPT CDM+baryons scheme (1F-cb)}

The second scheme is similar to the first one, but uses instead the CDM+baryons
linear power spectrum as input for the EdS-SPT loop
corrections~\cite{Saito:2008bp,Castorina:2015bma}:
\begin{equation}
    P_{\mm} = P_{\mm}^{\lin}
    + (1 - \fnu)^2 P_{\oneloop}^{\text{EdS-SPT}}(P_{\cb,\cb}^{\lin})
    + (1 - \fnu)^2 P_{\twoloop}^{\text{EdS-SPT}}(P_{\cb,\cb}^{\lin})\, .
\end{equation}
As the 1F scheme, this is reasonably straightforward to implement, but it
suffers from the same imprecisenesses. It will be labeled 1F-cb from now on.

\paragraph{External source scheme (1F-ext)}

The effect on the CDM and baryons due to the presence of massive neutrinos
enters through the coupling via gravitation, and one way to approximate this is
to use the linear neutrino transfer function in the Poisson
equation~\cite{Lesgourgues:2009am}. In our notation, this scheme (which we
label 1F-ext) amounts to using
\begin{equation}
    \Omega(k,\etaD) =
    \begin{pmatrix}
        0 & -1 \\
        - \frac{3}{2} \omegaoverfsq
        \xi(k,\etaD)
          & \frac{3}{2} \omegaoverfsq - 1
    \end{pmatrix}
\end{equation}
in the equations of motion for a one-component fluid
$\psi_a = (\delta_{\cb},-\theta_{\cb}/\mH f)$ with
\begin{equation}
    \xi(k,\etaD) = 1 - \fnu + \fnu
    \left(
        \frac{\delta_{\nu}(k,\etaD)}{\delta_{\cb}(k,\etaD)}
    \right)_{\lin}\, ,
\end{equation}
and $(\delta_{\nu}/\delta_{\cb})_{\lin}$ taken as external input. This scheme
is readily realized in our numerical setup for loop corrections in models with
time- and scale-dependent growth. By the same arguments as for the two-fluid
model (and to perform a comparison on equal footing), we use the 1F-ext scheme
in a hybrid setup: using the Boltzmann hierarchy (CLASS) up until a redshift
$\zmatch$ before non-linearities become important and matching onto the
one-component fluid. We use again $\zmatch = 25$, and take the linear transfer
function ratio $(\delta_{\nu}/\delta_{\cb})_{\lin}$ from CLASS.

Analogously to the two-fluid schemes, one needs to carefully choose the initial
conditions for the kernels in the 1F-ext scheme, in order to avoid transient
modes still being present at $\etaD = 0$. The prescription for initial
conditions described in the beginning of this section may equally well be
applied in the 1F-ext model: we evolve the kernel hierarchy from an early time
$\etaDasymp \ll \etaDmatch$ with $\Omega = \Omega(k,\etaDmatch)$ to ensure that
the kernels are fully settled in the growing mode solution at $\etaDmatch$.

\paragraph{Two-fluid scheme with adiabatic sound velocity (2F-ad)}

In this scheme the neutrino density contrast and velocity divergence are
included as a separate component, using the adiabatic approximation for the
neutrino sound velocity and neglecting the anisotropic stress, as discussed in
Sec.~\ref{sec:2fluid_lin} and at the beginning of this section.

\paragraph{Two-fluid scheme with exact effective sound velocity (2F)}

Finally, the 2F scheme is the one which best captures the neutrino dynamics and
its impact on the CDM+baryons fluid beyond the linear level, hence we use it as
the benchmark for the comparison. It incorporates the full scale-dependent
sound velocity as well as the impact of anisotropic stress on the linear
evolution matrix. The 2F scheme was described in detail in
Sec.~\ref{sec:2fluid_lin}, and the generalization to non-linear corrections was
laid out in the beginning of this section.

\subsection{Comparison}

\begin{figure}[t]
    \centering
    \includegraphics[width=\linewidth]{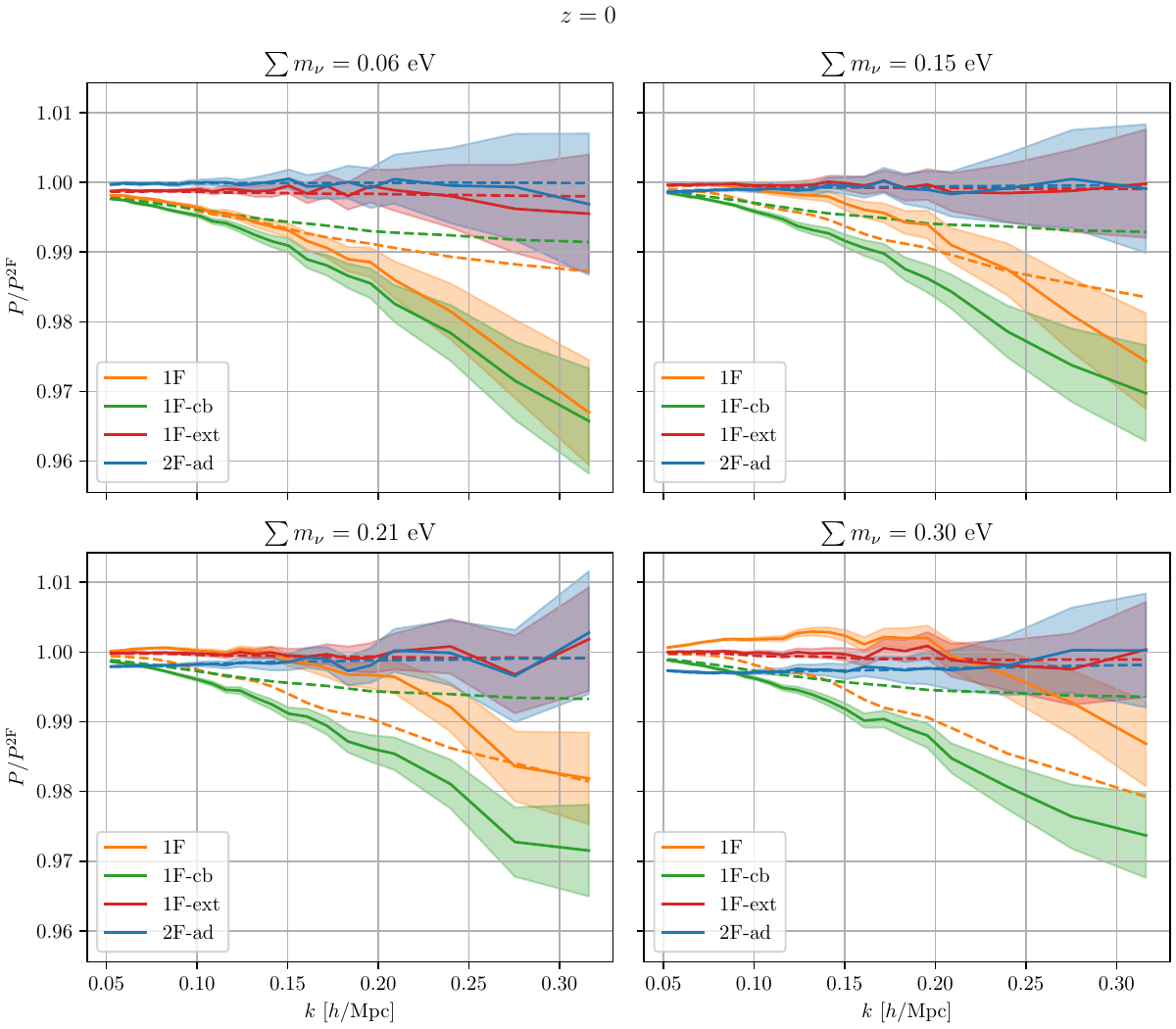}
    \caption{Fractional difference between different schemes and the 2F scheme
        of the total matter power spectrum $P_{\mm}$ at redshift $z = 0$ and
        with $\sum\mnu = 0.06, 0.15, 0.21, 0.3~\eV$. Dashed lines correspond to the 1-loop power
        spectrum $P = P^{\lin} + P^{\oneloop}$, while solid lines correspond to
        the 2-loop power spectrum $P = P^{\lin} + P^{\oneloop} + P^{\twoloop}$.
        The shaded areas indicate the numerical uncertainties of the 2-loop
    results (numerical uncertainty at 1-loop is invisible).
    Note that the apparent agreement of the 1F and 2F schemes at 2-loop for
    $\sum\mnu = 0.15, 0.21, 0.3~\eV$ and small $k$ is due to an accidental cancellation of
    inaccuracies in the individual 1- and 2-loop contributions that have opposite sign
    (see text for details and Fig.~\ref{fig:1F_vs_2F_L12}).
    Therefore, the total power spectrum cannot be regarded as a faithful measure
    of the precision of the 1F scheme (i.e.\ it performs worse than one would conclude
    from a comparison of the power spectrum only), in contrast to 1F-cb.
    }
    \label{fig:all_mnu005_L012}
\end{figure}

\begin{figure}[t]
    \centering
    \includegraphics[width=\linewidth]{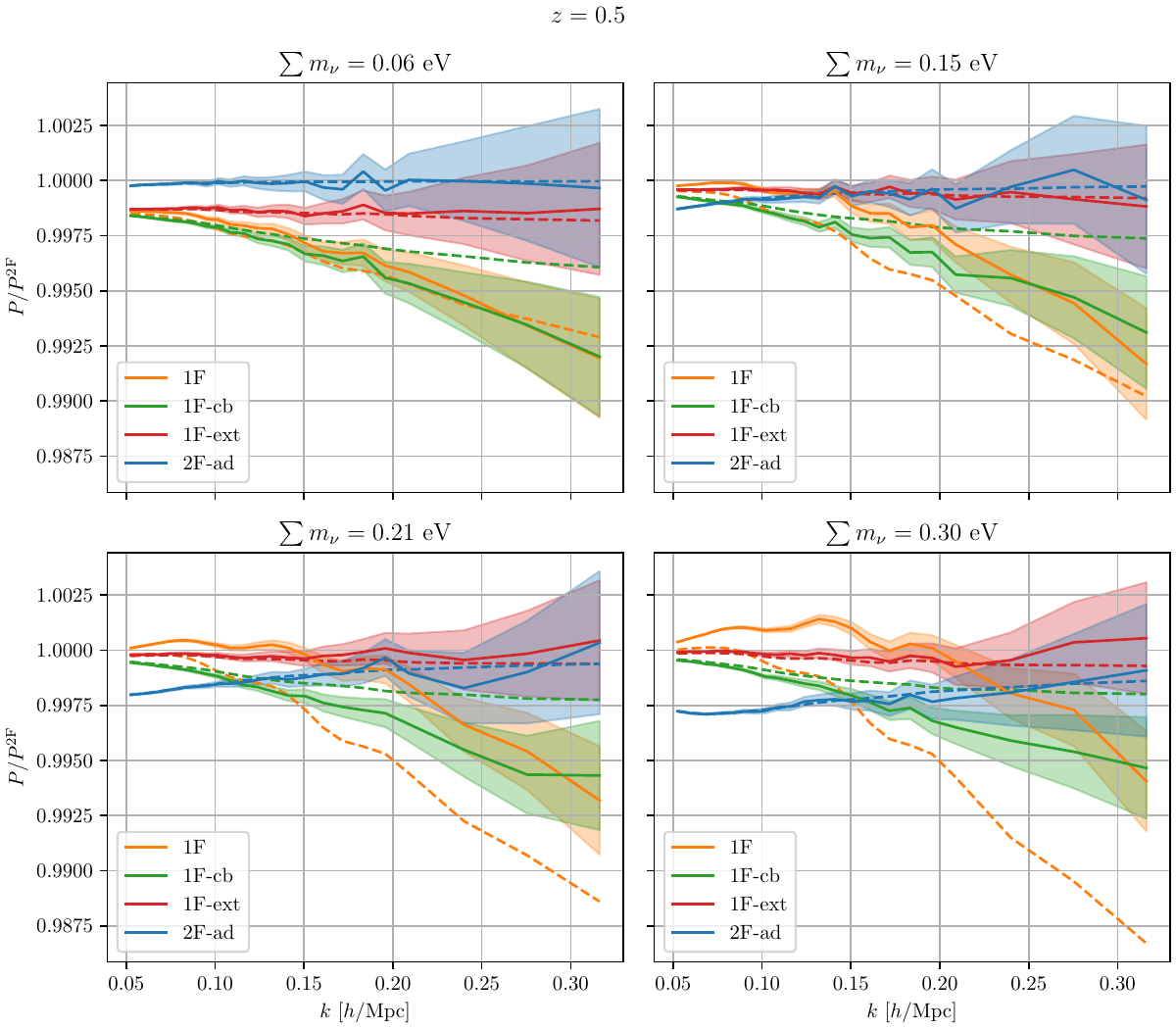}
    \caption{Same as Fig.~\ref{fig:all_mnu005_L012}, at redshift $z = 0.5$.}
    \label{fig:all_mnu005_z05_L012}
\end{figure}

In the following, we discuss the performance of the various approximation
schemes in capturing the effect of massive neutrinos on the non-linear power
spectrum. The results are summarized in Figs.~\ref{fig:all_mnu005_L012} and
\ref{fig:all_mnu005_z05_L012}, which show the non-linear corrections in the
different schemes normalized to the 2F scheme for various neutrino masses at
redshift $z=0$ and $z=0.5$, respectively. At large scales, $k \lesssim 0.1~h/\Mpc$,
the linear power spectrum dominates and all schemes largely agree. The tiny
discrepancies at the permille level arise due to the slight difference between
the linear two-fluid model and CLASS, as discussed in
Sec.~\ref{sec:2fluid_lin}. This could easily be corrected for by using the
linear power spectrum from CLASS also in the 2F and 2F-ad schemes. At smaller
scales, the non-linear corrections become important, and the approximations 1F
and 1F-cb perform progressively worse for increasing wavenumber. The 1F-ext and
2F-ad schemes on the other hand capture the time- and scale-dependence of the
dynamics more thoroughly, and remain in good agreement with 2F. A similar
behavior is seen at redshift $z = 0.5$, although to a lesser degree, since the
non-linear corrections are smaller compared to the linear power spectrum at
earlier times. We consider each scheme in detail in what follows.

\begin{figure}[t]
    \centering
    \includegraphics[width=\linewidth]{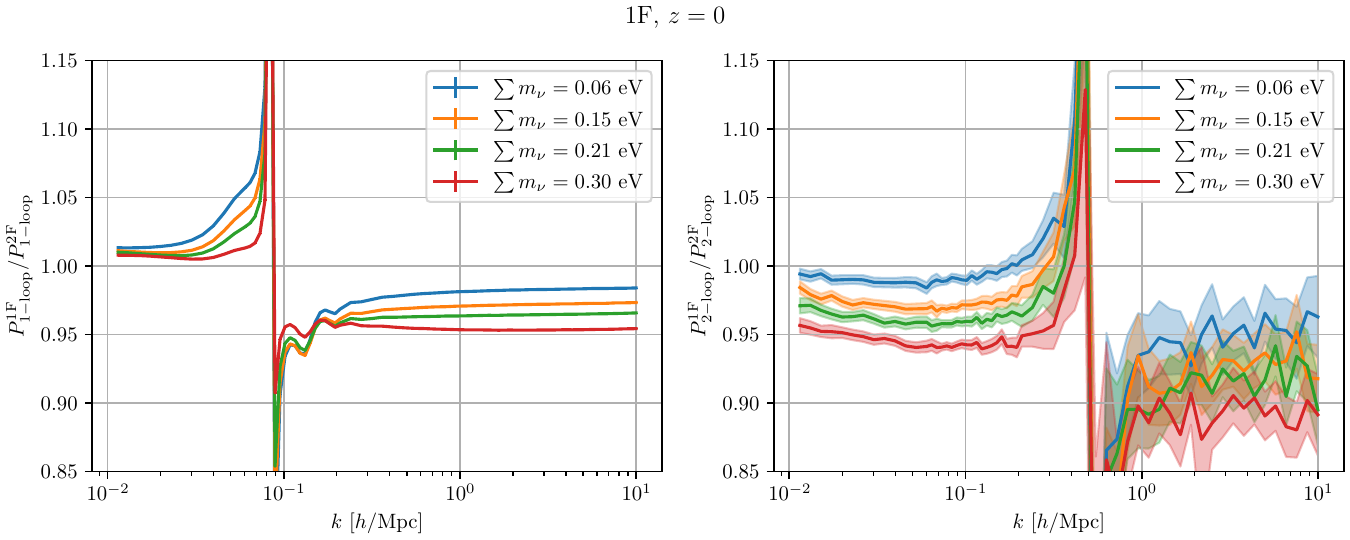}
    \caption{Non-linear corrections to the power spectrum for various neutrino
        masses at $z = 0$ computed in the 1F scheme, normalized to the 2F result.
        \emph{Left:} 1-loop correction. \emph{Right:} 2-loop correction. Shaded
        regions indicate uncertainty from numerical integration. At $k \simeq
        0.08~h/\Mpc$ and $k \simeq 0.5~h/\Mpc$ the 1- and 2-loop corrections cross
    zero, respectively, leading to large relative deviations.}
    \label{fig:1F_vs_2F_L12}
\end{figure}

\paragraph{EdS-SPT scheme (1F)}
The separate 1- and 2-loop corrections to the matter power spectrum in the 1F
scheme is plotted in Fig.~\ref{fig:1F_vs_2F_L12}. Even though small scales are
beyond the scope of perturbation theory, we note that the asymptotic behavior
for large $k$ can be understood in the following way: in this limit the
neutrino contribution to the total power spectrum becomes negligible so that
$P_{\mm} \simeq (1 - \fnu)^2 P_{\cb,\cb}$. In the 1F scheme, the loop
corrections to the CDM+baryons power spectrum is then computed with
$(1 -\fnu)^2 P_{\cb,\cb}$ as initial condition rather than $P_{\cb,\cb}$,
leading to a suppression by a factor of $2L\fnu$ compared to the 2F result on
these scales, with $L$ being the loop order. This is evident for the 1-loop
term at $k \gtrsim 1~h/\Mpc$ in Fig.~\ref{fig:1F_vs_2F_L12}. Due to the
coupling of Fourier modes, the suppression is to greater extent present across
scales in the 2-loop correction. For the highest neutrino masses, both the 1-
and 2-loop corrections are smaller than the corresponding terms in the 2F
scheme by about 5\% in the mildly non-linear regime. Since the 1- and 2-loop
terms are of the same order in this domain, but with opposite sign, the
deviations largely cancel out, leaving a remarkable agreement for the total power
spectrum, which is apparent in the lower plots of
Figs.~\ref{fig:all_mnu005_L012} and \ref{fig:all_mnu005_z05_L012}.
Due to the different dependence of the 1- and
2-loop corrections on redshift and other cosmological parameters such as $A_s$,
this cancellation should be considered as coincidental.

\begin{figure}[t]
    \centering
    \includegraphics[width=\linewidth]{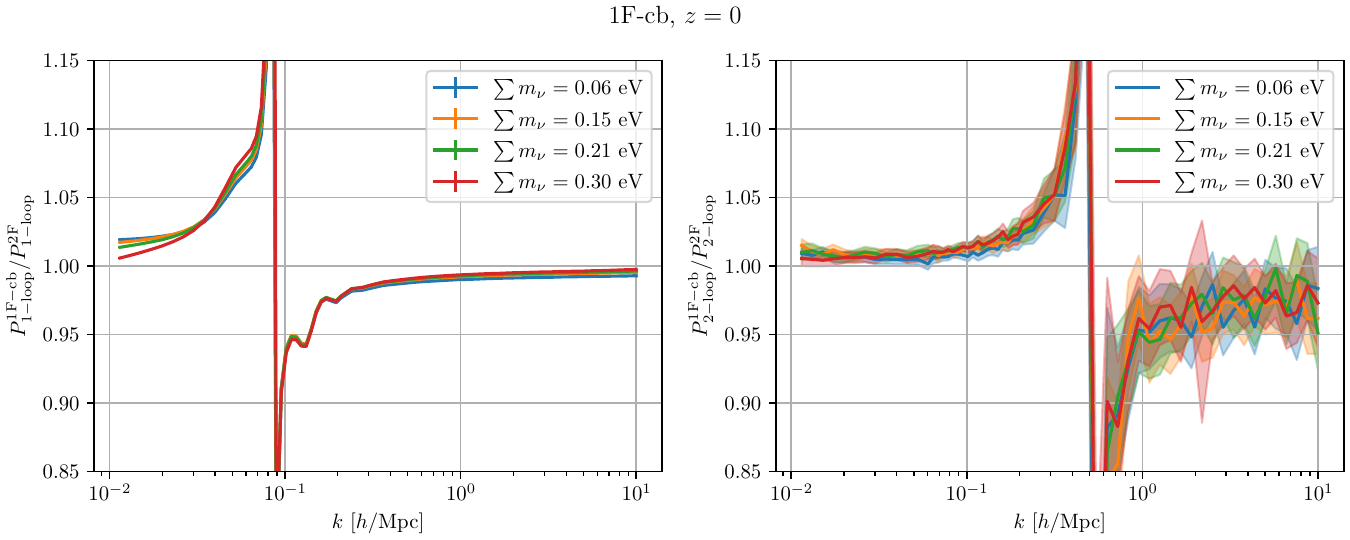}
    \caption{Same as Fig.~\ref{fig:1F_vs_2F_L12} with the 1F-cb scheme
    normalized to 2F.}
    \label{fig:1F-cb_vs_2F_L12}
\end{figure}

\paragraph{EdS-SPT CDM+baryons scheme (1F-cb)}

By taking the CDM+baryons power spectrum as input to the EdS-SPT loop
evaluation, the 1F-cb scheme corrects for the $\sim 2L\fnu$ suppression in the
1F scheme. This is seen in Fig.~\ref{fig:1F-cb_vs_2F_L12}, which displays the
separate 1- and 2-loop corrections to the power spectrum in the 1F-cb scheme,
normalized to the 2F scheme. Given our knowledge of the deviation in the
non-linear corrections due to the departure from EdS from
Sec.~\ref{sec:eds_approx} and comparing to Fig.~\ref{fig:lcdm_vs_eds_L12} in
particular, we conclude that the main difference between the 1F-cb and 2F
results comes from the EdS approximation in 1F-cb. The deviation due to the
neglection of the neutrino-neutrino and CDM+baryons-neutrino power spectra
is only about 0.1\%. Note that the 1F-cb scheme appears to perform worse than
the 1F scheme in Figs.~\ref{fig:all_mnu005_L012} and
\ref{fig:all_mnu005_z05_L012}, however this is only due to the accidental
cancellation between inaccuracies of the 1- and 2-loop terms in the 1F
scheme.

\paragraph{External source scheme (1F-ext)}
With a more proper treatment of the time- and scale-dependence of the dynamics,
the external source scheme yields results that are unwaveringly in agreement
with the 2F scheme. There is a small difference at the permille level for the
smallest neutrino mass, but this could equally well be attributed to the
two-fluid model: it differs about a permille to the CLASS solution at the
linear level, and this error plausibly propagates to higher orders. We note
that spurious dipole contributions may appear in the $k\to 0$ limit of the
power spectrum in the 1F-ext scheme, due to momentum conservation being
slightly broken~\cite{Blas:2014}. Moreover, the numerical complexity of the
1F-ext scheme is comparable to the two-fluid setup, such that in practice
there is little advantage over using 2F or 2F-ad.

\paragraph{Two-fluid scheme with adiabatic sound velocity (2F-ad)}
Lastly, the 2F-ad scheme also agrees reasonably well with the 2F scheme in the
linear and mildly non-linear regimes. Even though the neutrino transfer
functions are underestimated by $\mathcal{O}(10\%)$ at the linear level, this
has negligible impact on the total matter power spectrum for the lowest
neutrino mass. Increasing the neutrino mass, the error in the neutrino sector
influences the CDM+baryons to greater extent via the coupling through gravity,
leading to e.g.\ 0.3\% deviation in the mildly non-linear regime for
$\sum\mnu = 0.3~\eV$ compared to the 2F scheme.

In summary, we find that the external source scheme 1F-ext best emulates the
results of the 2F scheme. Of the simplest schemes 1F and 1F-cb, 1F-cb yields
the best results for the separate loop corrections, with the main source of
error coming from the EdS approximation. Even though the 1F scheme appears to
perform better when the linear and non-linear contributions are added, this is
due to a cancellation of errors, which is regarded as coincidental and moreover
not expected to necessarily occur when e.g.\ including redshift space
distortion effects or effective field theory corrections.

\section{Conclusions}
\label{sec:conclusions}

In this work, we have set up a framework that captures the effect of time-
and scale-dependent growth on non-linear corrections to the power spectrum.
This is achieved by extending SPT to a generic form that describes a
multi-component fluid with the linear evolution characterized by a time- and scale-dependent
matrix $\Omega(k,\etaD)$. The framework can be applied to a wide range
of extended cosmological models, and we use it in particular to examine the
reliability of the EdS approximation and the effect of massive neutrinos on
loop corrections to the matter power spectrum. As in SPT, we expand the density contrast
and velocity divergence in terms of the linear density contrast, but due to the
dependence on time and scale, the resulting equations of motion for the kernels
cannot be solved analytically as is possible in the EdS-SPT case. Instead we
solve the hierarchical differential equations for the kernels
(Eq.~\eqref{eq:eom_kernels}) numerically, for each configuration of wavenumbers
needed by the Monte-Carlo loop integration that yields the non-linear
corrections to the power spectrum in the perturbative expansion.

Our first application of the numerical framework is investigating the validity
of the commonly used EdS approximation for the matter power spectrum at 2-loop.
We checked that at 1-loop, we reproduce the results previously derived in the
literature: the EdS approximation is accurate to less than a percent at
redshift $z = 0$ in the mildly non-linear regime. The deviation increases to
more than 1\% at $k = 0.2~h/\Mpc$ when adding the 2-loop term. At redshift
$z = 0.5$, the error introduced by the EdS approximation is less, because
$\Omega_m/f^2$ is closer to unity and the non-linear corrections are suppressed
to greater extent compared to the linear power spectrum.

In order to be able to calculate non-linear corrections to the power spectrum
in the presence of massive neutrinos, we employ a refined version of the
two-fluid model of Ref.~\cite{Blas:2014}. It utilizes the fact that after the
non-relativistic transition but before non-linearities become important, the
lower and higher moments of the Boltzmann hierarchy decouple, so the neutrinos
may be described by a fluid with an effective sound velocity. Thereby the
CDM+baryons and neutrinos are modeled as separate components of a two-component
fluid, with coupling through gravity. We demonstrate two ways to model the
effective neutrino sound velocity: using an adiabatic approximation and
integrating the linearized Boltzmann equation to obtain sound velocity and
anisotropic stress. Comparing the latter with the Boltzmann solver CLASS, we
obtain permille and percent agreement for the CDM+baryons and neutrinos,
respectively.

The two-fluid setup for massive neutrino cosmologies is readily realized
in our numerical framework for investigating the impact of neutrino free-streaming
on weakly non-linear scales.
We compare the 1- and 2-loop matter power spectrum with full time- and scale-dependence
to various simplified approximations that treat the neutrinos linearly. The approximation scheme that
best captures the neutrino backreaction on the CDM+baryons is the external
source scheme (1F-ext), which we find to agree excellently with the two-fluid
model. However, the computational complexity for 1F-ext is comparable to the
full two-fluid treatment.
Approximating the loop corrections by $P^{\text{EdS-SPT}}(P_{\mm}^{\lin})$ (1F scheme),
we find deviations of several percent for the individual 1- and 2-loop corrections
compared to the two-fluid model. While the mismatch largely cancels when adding
1- and 2-loop contributions, this cancellation among terms of different order
in the perturbative expansion should be considered as purely accidental.
When using only the CDM+baryon power spectrum as input to compute loop
corrections, $P^{\text{EdS-SPT}}(P_{\cb,\cb}^{\lin})$ (1F-cb scheme), the 1- and
2-loop corrections are in better agreement with the two-fluid treatment. The total
matter power spectrum deviates
by about $1\%$ ($z = 0$) and $0.25\%$ ($z = 0.5$) from the two-fluid model at
$k = 0.15~h/\Mpc$, with the predominant error coming from the EdS approximation.

Given the versatility of our numerical framework, it is suitable for precision
analyses incorporating relativistic effects, small-scale effective stress-tensor corrections
and redshift space distortions in addition to describing extended cosmological
models in the future.

\acknowledgments{%
This work was supported by the DFG Collaborative Research Institution
\emph{Neutrinos and Dark Matter in Astro- and Particle Physics}
(\href{www.sfb1258.de}{SFB 1258}).
}

\appendix

\section{Validation with analytic 1-loop kernels}
\label{app:ana_kernels}

In this appendix, we compare the power spectrum at 1-loop obtained using
numerically evolved kernels and using analytic expressions for the kernels
when the EdS approximation is relaxed. Expressions for the second- and
third-order generalized kernels with time-dependence have been derived in the
past~%
\cite{Bernardeau:1993qu,Takahashi:2008yk,Sefusatti:2011cm,Fasiello:2016qpn,Donath:2020abv},
and we repeat them here for convenience. At second order, the kernels may be
written in such a way that the time-dependence is contained in the
angle-averaged kernels $\nu_2$ and $\mu_2$,
\begin{align}
    F_{2}(\vq_1,\vq_2;\etaD) & =
    \left(
        - \frac{1}{2}
        + \frac{3}{4} \nu_2(\etaD)
    \right) \alphas (\vq_1,\vq_2)
    +
    \left(
        \frac{3}{2}
        - \frac{3}{4} \nu_2(\etaD)
    \right) \beta (\vq_1,\vq_2)
    \label{eq:gen_kernel_F2}\,, \\
    G_{2}(\vq_1,\vq_2;\etaD) & =
    \left(
        - \frac{1}{2}
        + \frac{3}{4} \mu_2(\etaD)
    \right) \alphas (\vq_1,\vq_2)
    +
    \left(
        \frac{3}{2}
        - \frac{3}{4} \mu_2(\etaD)
    \right) \beta(\vq_1,\vq_2) \, ,
    \label{eq:gen_kernel_G2}
\end{align}
where $\alphas(\vq_1,\vq_2) = (\alpha(\vq_1,\vq_2) + \alpha(\vq_2,\vq_1))/2$.
Note that an overall factor $\e^{n\etaD}$ was extracted from the kernels above
compared to those defined in Eq.~\eqref{eq:Fan}: in our notation
from Sec.~\ref{sec:extension} we have
$F_1^{(2)}(\etaD) = \e^{2\etaD}\,F_2(\etaD)$ and
$F_2^{(2)}(\etaD) = \e^{2\etaD}\,G_2(\etaD)$.
The angle-averaged kernels are momentum-independent and satisfy the following
equations of motion:
\begin{align}
    \partial_{\etaD}
    \begin{pmatrix}
        \nu_n \\
        \mu_n
    \end{pmatrix}_a
    + n
    \begin{pmatrix}
        \nu_n \\
        \mu_n
    \end{pmatrix}_a
    + \Omega_{ab}(\etaD)
    \begin{pmatrix}
        \nu_n \\
        \mu_n
    \end{pmatrix}_b
    =
    \sum_{m=1}^{n-1} {n \choose m}
    \begin{pmatrix}
        \mu_m \, \nu_{n-m} \\
        \frac{1}{3} \, \mu_m \, \mu_{n-m}
    \end{pmatrix}_a \, ,
\label{eq:angle_averaged_kernels}
\end{align}
where $\Omega_{ab}$ was defined in Eq.~\eqref{eq:1fluid_omega_mat}. Given the
linear growing mode solution $\nu_1 = \mu_1 = 1$, the equation of motion for
$\nu_2$ and $\mu_2$ may easily be integrated numerically. Due to the
factorization of the time- and scale-dependence in
Eqs.~\eqref{eq:gen_kernel_F2} and \eqref{eq:gen_kernel_G2}, one only needs to
do this once given a cosmological model. In contrast, the numerical setup
described in Sec.~\ref{sec:extension} necessitates solving a similar set of
differential equations separately for each wavenumber configuration required
by the Monte-Carlo loop integration. Factorizing the time- and scale-dependence
of generalized kernels becomes progressively more difficult at higher orders in
perturbation theory, however; as far as we are aware there is no derivation of
this beyond third order. We note finally that in the EdS limit, $\nu_2 = 34/21$
and $\mu_2 = 26/21$, which inserted in Eqs.~\eqref{eq:gen_kernel_F2} and
\eqref{eq:gen_kernel_G2} reproduces the EdS-SPT kernels.

At third order, the time-dependence of the $F_3$ kernel can be extracted into
the parameters $\nu_2$, $\nu_3$ and
$\lambda_3$~\cite{Bernardeau:2001qr, Bernardeau:1993qu},%
\footnote{A similar expression for $F3$ is derived in
Ref.~\cite{Fasiello:2016qpn}, which is related to Eq.~\eqref{eq:gen_kernel_F3} by
$\lambda_1 = 8/9 - \nu_2 + \nu_3/2 + 2\lambda_3/9$ and
$\lambda_2 = 13/18 - 3\nu_2/4 + 3\nu_3/8 + \lambda_3/18$.}
\begin{align}
    F_3(\vq_1, \vq_2, \vq_3; \etaD) & =
    \mathcal{R}_1(\vq_1, \vq_2, \vq_3)
    + \nu_2(\etaD) \mathcal{R}_2(\vq_1, \vq_2, \vq_3)
    + \nu_3(\etaD) \mathcal{R}_3(\vq_1, \vq_2, \vq_3)
    \nonumber \\
    & \phantom{ = }
    + \lambda_3(\etaD) \mathcal{R}_4(\vq_1, \vq_2, \vq_3) \, ,
    \label{eq:gen_kernel_F3}
\end{align}
where $\lambda_3(\etaD)$ is a function satisfying
\begin{equation}
    \left(
        \partial_{\ln a}^2
        +
        \left(
            1 + \frac{\dd \ln \mH}{\dd \ln a}
        \right)
        \partial_{\ln a}
        - \frac{3}{2} \Omega_m
    \right)
    (\lambda_3(a)D^3(a))
    = \frac{3}{2} \Omega_m D^3(a) \, .
    \label{eq:lambda_3}
\end{equation}
Using the notations $\alpha_{ij,k} = \alpha(\vq_i + \vq_j,\vq_k)$ and
$\alpha_{i,jk} = \alpha(\vq_i,\vq_j + \vq_k)$, the momentum-dependent functions
in the expression for $F_3$ are
\begin{subequations}
    \begin{align}
        \mathcal{R}_{1}&=
        \left(
            \frac{2}{3} \alphas_{3,12} + \frac{1}{3} \beta_{3,12}
        \right)
        \alpha_{1,2}+
        \left(
            \frac{1}{6}\alpha_{3,12}
            - \frac{1}{2} \alphas_{3,12}
            - \frac{5}{2} \beta_{3,12}
        \right)
        \left(
            \alphas_{1,2} - \beta_{1,2}
        \right), \\
        \mathcal{R}_{2} &=
        \frac{3}{4}
        \left(
            3 \beta_{3,12} - \alphas_{3,12}
        \right)
        \left(
            \alphas_{1,2} - \beta_{1,2}
        \right), \\
        \mathcal{R}_{3}&=
        \frac{3}{8}
        \left(
            \alphas_{3,12} - \beta_{3,12}
        \right)
        \left(
            \alphas_{1,2} - \beta_{1,2}
        \right), \\
        \mathcal{R}_{4}&=
        \frac{2}{3}
        \left(
            \alphas_{3,12} - \beta_{3,12}
        \right)
        \alpha_{1,2} -
        \left(
            \frac{1}{3} \alpha_{3,12}
            + \frac{1}{2} \alphas_{3,12}
            - \frac{1}{2} \beta_{3,12}
        \right)
        \left(
            \alphas_{1,2} - \beta_{1,2}
        \right) \, .
    \end{align}
\end{subequations}
We symmetrize the above expressions with respect to permutations of $\vq_1$,
$\vq_2$ and $\vq_3$ in order to utilize the loop integration algorithm
described in Sec.~\ref{sec:extension}.

\begin{figure}[t]
    \centering
    \includegraphics[width=\linewidth]{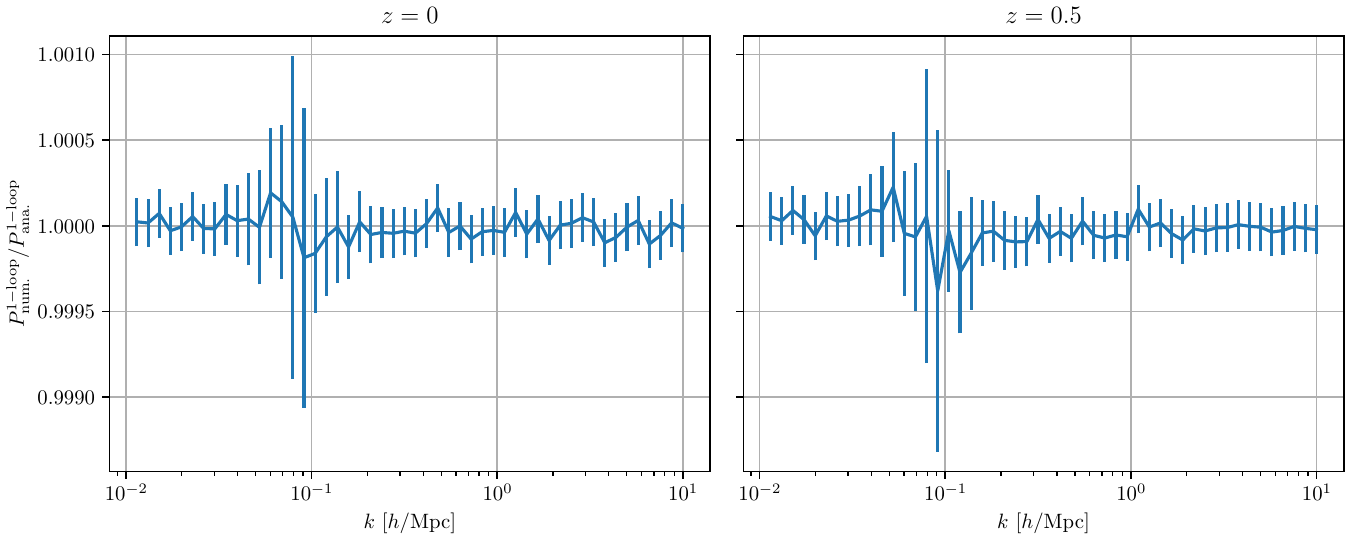}
    \caption{Ratio of the 1-loop correction to the matter power spectrum
    computed from numerically evolved kernels and computed from the analytic
    kernel expressions. The $\nu_2$, $\nu_3$ and $\lambda_3$ parameters used
    are listed in Table~\ref{tab:gen_kernels_parameters}. \emph{Left:} $z = 0$.
    \emph{Right:} $z = 0.5$. The error bars correspond to uncertainty from the
    numerical integration.}
    \label{fig:gen_kernels}
\end{figure}

\begin{table}[t]
	\begin{center}
        \caption{Parameters used in expressions for generalized kernels. The
        constant values in EdS are listed as reference.}
		\label{tab:gen_kernels_parameters}
		\vspace{0.5cm}
		\begin{tabular}{lllll}
			\hline
             &  $\nu_2$  & $\mu_2$ & $\nu_3$ & $\lambda_3$ \\
			\hline
            EdS                & 34/21     & 26/21       & 682/189   & 1/6 \\
            $\LCDM$, $z = 0$   & 1.6215126 & 1.2510105 & 3.6228998 & 0.1696433 \\
            $\LCDM$, $z = 0.5$ & 1.6201647 & 1.2434840 & 3.6149456 & 0.1679576 \\
			\hline
		\end{tabular}
	\end{center}
\end{table}

In Fig.~\ref{fig:gen_kernels}, we compare the power spectrum at 1-loop using
kernels computed from Eqs.~\eqref{eq:gen_kernel_F2}, \eqref{eq:gen_kernel_G2}
and \eqref{eq:gen_kernel_F3} with that computed by solving
Eq.~\eqref{eq:eom_kernels} (with $\Omega$ given in
Eq.~\eqref{eq:1fluid_omega_mat}) numerically.
The time-dependent parameters in the generalized kernels are evaluated
numerically by solving Eqs.~\eqref{eq:angle_averaged_kernels} and
\eqref{eq:lambda_3}, yielding the values listed in
Table~\ref{tab:gen_kernels_parameters}. We find excellent agreement between the
power spectra from the analytic kernels and from the numerically evolved ones.

\section{Numerical checks of convergence}
\label{app:check}

We performed various numerical checks to ensure the stability of the algorithm.
First of all, we checked that when using the EdS-SPT recursion relation for the
kernels (rather than evolving them numerically), we obtain agreement between
two independent implementations computing the 2-loop power spectrum.  Secondly,
when evolving the kernels numerically according to Eq.~\eqref{eq:eom_kernels}
using the constant evolution matrix $\Omega = \Omega^{\mathrm{EdS}}$, we
confirm that we recover the results of the EdS-SPT recursion relation. Finally,
we checked that changing various parameters in the numerical setup does not
change the results beyond the numerical uncertainty. We discuss the last point
in detail in the following.

\begin{figure}[t]
    \centering
    \includegraphics[width=\linewidth]{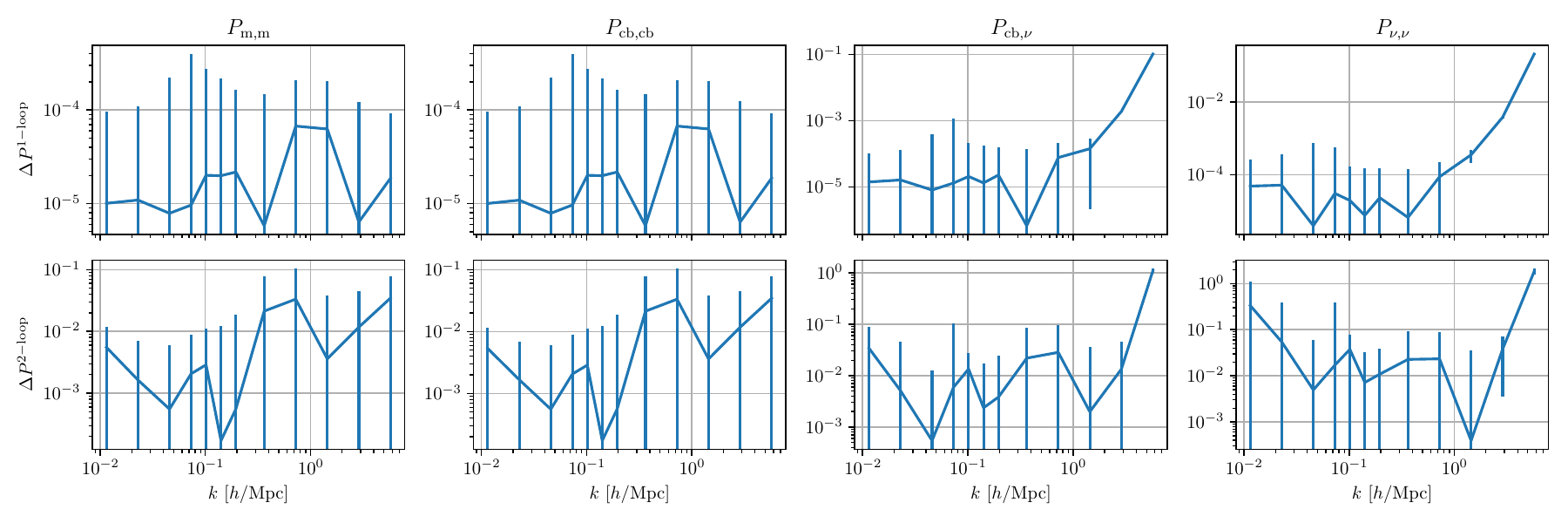}
    \caption{Relative difference of the various contributions to the 1- and
        2-loop corrections to the power spectrum when changing the number of
        grid points between the initial and final times, i.e.~%
        $\Delta P = | P^{N=125}/P^{N=100} - 1 |$. Neutrino mass:
        $\sum m_{\nu} = 0.15~\eV$. The top and bottom rows
        correspond to the 1- and 2-loop correction, respectively. The columns
        display from left to right: $P_{\mm}$, $P_{\cb,\cb}$, $P_{\cb,\nu}$ and
    $P_{\nu,\nu}$.}
    \label{fig:2fluid_N}
\end{figure}

As described in Sec.~\ref{sec:extension}, for each wavenumber configuration the
kernels computed by Eq.~\eqref{eq:eom_kernels} are temporarily stored on a grid
between the initial and final times, so that they can potentially be used in
the RHS of the same equation when solving for a higher order kernel. We used
$N=100$ as grid size in both the analysis of the departure from EdS in
Sec.~\ref{sec:eds_approx} and the analysis of the effect of massive neutrinos
on the matter power spectrum in Sec.~\ref{sec:2fluid_nonlin}. For the one-fluid
implementation used in Sec.~\ref{sec:eds_approx}, increasing the grid size to
$N=150$ does not alter the results beyond the numerical uncertainty. Likewise,
setting $N=125$ in the two-fluid model for massive neutrinos (scheme 2F)
impacts the results insignificantly. In Fig.~\ref{fig:2fluid_N} we show the
relative difference between results obtained with $N=125$ and $N=100$ for a
subset of wavenumbers and for all the 1- and 2-loop corrections to the total
power spectrum in a cosmology with neutrino mass $\sum\mnu = 0.15~\eV$. At
small scales, there is a relative deviation of order $\mathcal{O}(1)$ in
$P_{\cb,\nu}$ and $P_{\nu,\nu}$. However, $P_{\cb,\nu}$ and $P_{\nu,\nu}$ are
suppressed by several orders of magnitude compared to $P_{\cb,\cb}$ in that
domain, and the total power spectrum receives no significant change from
altering $N$. In addition, small scales are anyway beyond the reach of
perturbation theory; in the linear and mildly non-linear regime none of the
contributions differ beyond the numerical uncertainties when changing $N$.
We perform the same check for the other neutrino masses considered in this
work, also finding deviations that are smaller than the numerical uncertainty.

\begin{figure}[t]
    \centering
    \includegraphics[width=\linewidth]{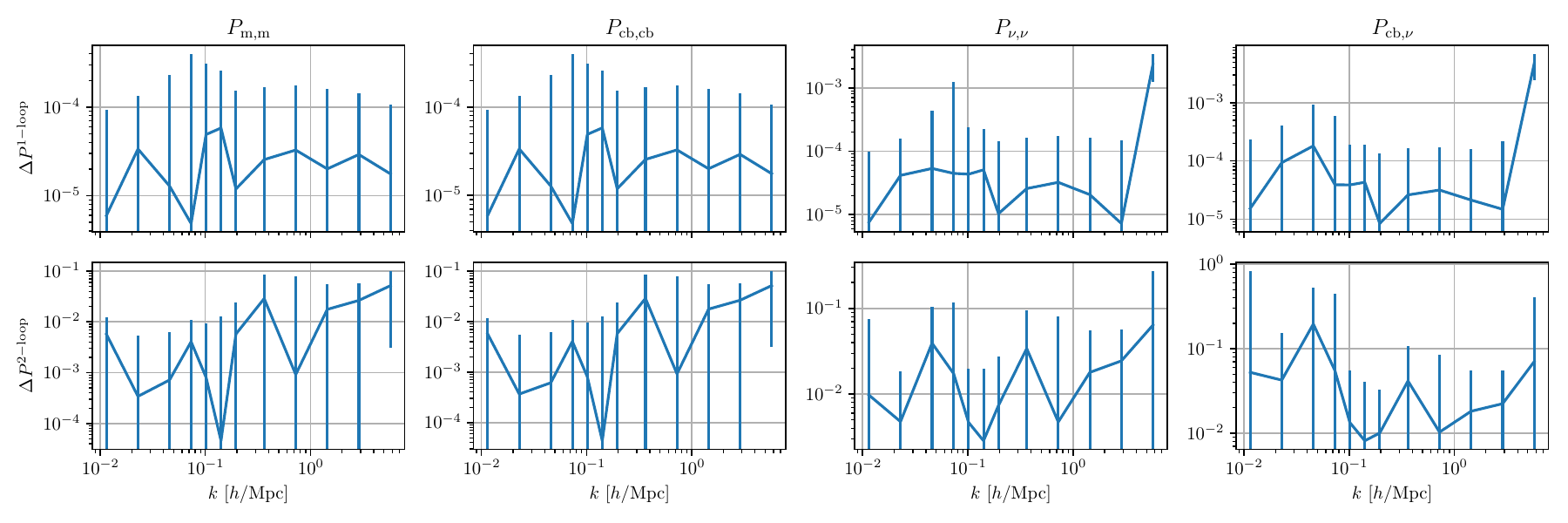}
    \caption{Relative difference of the various contributions to the 1- and
        2-loop corrections to the power spectrum when adjusting the lower
        integration cutoff in the loop integration, i.e.~%
        $\Delta P = | P(q_{\mathrm{min}} = 1.3\cdot 10^{-5}~h/\Mpc)/P(q_{\mathrm{min}} = 10^{-4}~h/\Mpc) - 1|$.
        Neutrino mass: $\sum m_{\nu} = 0.15~\eV$. The top and bottom rows
        correspond to the 1- and 2-loop correction, respectively. The columns
        display from left to right: $P_{\mm}$, $P_{\cb,\cb}$, $P_{\cb,\nu}$ and
    $P_{\nu,\nu}$.}
    \label{fig:2fluid_qmin}
\end{figure}

\begin{figure}[t]
    \centering
    \includegraphics[width=\linewidth]{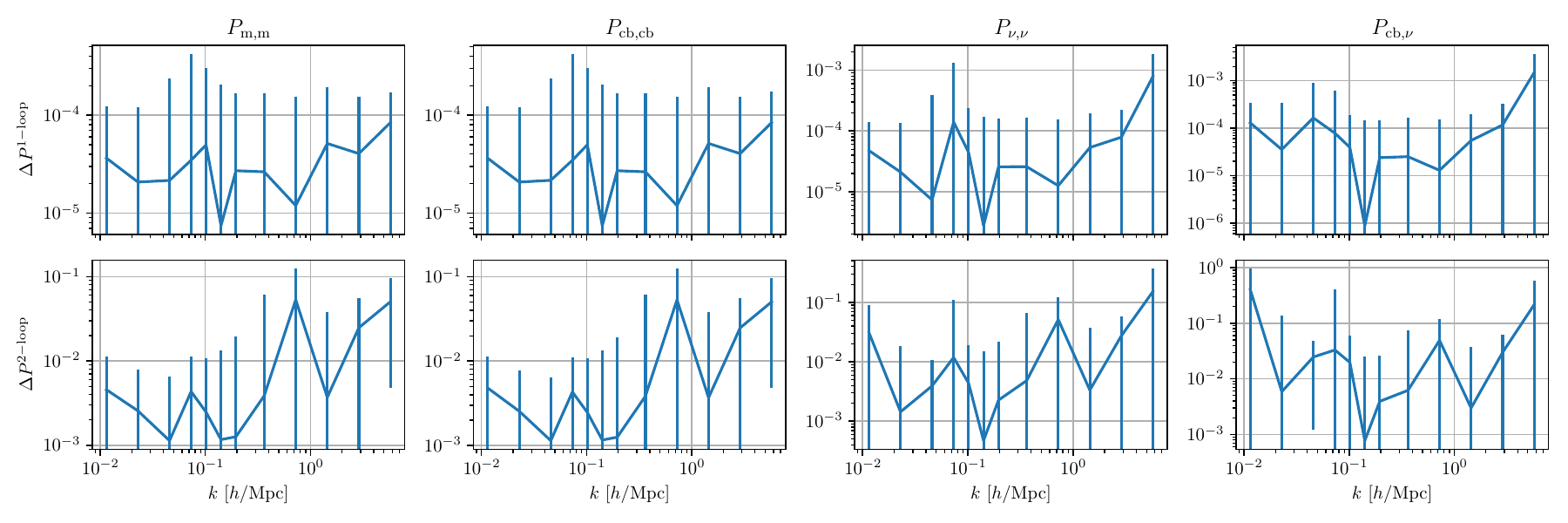}
    \caption{Relative difference of the various contributions to the 1- and
        2-loop corrections to the power spectrum when adjusting the upper
        integration cutoff in the loop integration, i.e.~%
        $\Delta P = | P(q_{\mathrm{max}} = 70~h/\Mpc)/P(q_{\mathrm{max}} = 65~h/\Mpc) - 1|$.
        Neutrino mass: $\sum m_{\nu} = 0.15~\eV$. The top and bottom rows
        correspond to the 1- and 2-loop correction, respectively. The columns
        display from left to right: $P_{\mm}$, $P_{\cb,\cb}$, $P_{\cb,\nu}$ and
    $P_{\nu,\nu}$.}
    \label{fig:2fluid_qmax}
\end{figure}

We used throughout $q_{\mathrm{min}} = 10^{-4}~h/\Mpc$ and
$q_{\mathrm{max}} = 65~h/\Mpc$ as integration limits in the Monte-Carlo
integration. This domain covers the scales in which the linear power spectrum
is largest and hence also the regions where the integrand receives its dominant
contributions. Indeed, beyond the chosen integration limits the linear power
spectrum is suppressed by several orders of magnitude compared to the peak.
Moreover, due to the cancellations of large terms in the integrand when the
external wavenumber is large, one cannot extend the upper limit too far: the
terms in question grow with increasing separation between the momenta (both
loop and external) and at some point the cancellation might be slightly spoiled
by machine rounding error. We find that using an upper integration limit around
$q_{\mathrm{max}} = 65$, the one-fluid calculation used in
Sec.~\ref{sec:eds_approx} is stable against modest changes of
$q_{\mathrm{max}}$. In addition, decreasing the lower integration limit to
$q_{\mathrm{min}} = 1.3\cdot 10^{-5}$ does not alter the results beyond the
numerical error bars. A similar exercise was performed for the two-fluid model
with massive neutrinos: in Fig.~\ref{fig:2fluid_qmin} we show the relative
difference introduced in the 1- and 2-loop corrections when decreasing the
lower integration limit to $1.3\cdot 10^{-5}$ (neutrino mass $\sum\mnu =
0.15~\eV$). Corresponding plots for the relative difference between results
with $q_{\mathrm{max}} = 65$ and $q_{\mathrm{max}} = 70$ are shown in
Fig.~\ref{fig:2fluid_qmax}. We find agreement within the numerical uncertainty
both when changing the lower and upper integration cutoffs, except for a few of
the values on small scales, which are anyway far beyond the scope of
perturbation theory. Similar agreement is found for the other neutrino masses
used in this work.

Let us finally comment on the computational complexity of the schemes considered
in this work. The main additional running time compared to EdS-SPT kernels is
required for solving the ordinary differential equations for the time-dependent
kernels. For the fiducial choice of parameters discussed above, the 1F-ext scheme
has an overhead of around a factor $100$ compared to 1F, and 2F as well as 2F-ad
of $\sim 500$. For 2F, the exact sound velocity and anisotropic stress need to be
computed in addition. However, they can be pre-computed for a sufficiently dense
grid in wavenumber. The running time of 1F and 1F-cb is identical, since both schemes
are based on EdS-SPT kernels. In this work we focus on achieving maximal precision, in order
to quantify the accuracy of the computationally more efficient schemes 1F and 1F-cb.
Nevertheless, by optimizing the numerical settings used in the 1F-ext and 2F/2F-ad
schemes, we expect that their running time can be reduced.

\providecommand{\href}[2]{#2}\begingroup\raggedright\endgroup

\end{document}